\documentclass[preprint,12pt]{elsarticle}

\newcommand{\equref}[1]{(\ref{#1})}
\graphicspath{{images/}}

\usepackage{titlesec}
\usepackage{amssymb}
\usepackage{amsmath}
\usepackage{courier}
\usepackage{algorithm}
\usepackage{algpseudocode}
\usepackage{booktabs}
\usepackage{t1enc}
\usepackage[utf8]{inputenc}
\algnewcommand\algorithmicforeach{\textbf{for each}}
\algdef{S}[FOR]{ForEach}[1]{\algorithmicforeach\ #1\ \algorithmicdo}
\usepackage{listings}
\usepackage[many]{tcolorbox}
\newtcbtheorem[]{example}{}{
  breakable,
  enhanced,
  colback=yellow!5,
  colframe=blue!40!gray,
  fonttitle=\bfseries}{x}

\biboptions{sort&compress}

\newcounter{bla}

\journal{Computer Physics Communications}

\newcommand{\customfigure}[3]{\begin{figure}[H]%
\includegraphics[scale=#1]{#2.pdf}%
\centering%
\caption{#3}\label{fig:#2}%
\end{figure}}

\definecolor{nauticle_green}{RGB}{91,122,0}
\definecolor{nauticle_blue}{RGB}{68,84,108}

\definecolor{revcolor}{RGB}{0,0,0} 
\newcommand{\revised}[1]{\color{revcolor}{#1}\color{black}{}}
\newcommand\YAMLcolonstyle{\color{nauticle_blue}\mdseries}
\newcommand\YAMLkeystyle{\color{nauticle_blue}\bfseries}
\newcommand\YAMLvaluestyle{\color{nauticle_green}\mdseries}

\makeatletter

\newcommand\language@yaml{yaml}

\expandafter\expandafter\expandafter\lstdefinelanguage
\expandafter{\language@yaml}
{
  keywordstyle=\color{darkgray}\bfseries,
  basicstyle=\YAMLkeystyle\footnotesize\ttfamily,                                 
  sensitive=false,
  comment=[l]{\#},
  morecomment=[s]{/*}{*/},
  commentstyle=\color{gray}\ttfamily,
  stringstyle=\YAMLvaluestyle\ttfamily,
  moredelim=[l][\color{orange}]{\&},
  moredelim=**[il][\YAMLcolonstyle{:}\YAMLvaluestyle]{:},   
  morestring=[b]',
  morestring=[b]",
  literate =    {---}{{\ProcessThreeDashes}}3
                {>}{{\textcolor{nauticle_green}\textgreater}}1     
                {|}{{\textcolor{nauticle_green}\textbar}}1 
                {\ -\ }{{\mdseries\ -\ }}3,
}
\lstdefinestyle{CPP} {language=C++}
\lstdefinestyle{YAML} {language=yaml}

\lst@AddToHook{EveryLine}{\ifx\lst@language\language@yaml\YAMLkeystyle\fi}
\makeatother

\newcommand\ProcessThreeDashes{\llap{\color{cyan}\mdseries-{-}-}}

\begin{document}
\begin{frontmatter}



\title{Nauticle: a general-purpose particle-based simulation tool }


\author[]{Balázs Tóth\corref{author}}

\cortext[author] {Corresponding author.\\\textit{E-mail address:} toth.balazs@epito.bme.hu}
\address{Department of Hydraulic and Water Resources Engineering, Budapest University of Technology and Economics, Műegyetem rkp. 3., Budapest, H-1111, HUNGARY}

\begin{abstract}
\revised{Nauticle is a general-purpose simulation tool for the flexible and highly configurable application of particle-based methods of either discrete or continuum phenomena. It is presented that Nauticle has three distinct layers for users and developers, then the top two layers are discussed in detail. The paper introduces the Symbolic Form Language (SFL) of Nauticle, which facilitates the formulation of user-defined numerical models at the top level in text-based configuration files and provides simple application examples of use. On the other hand, at the intermediate level, it is shown that the SFL can be intuitively extended with new particle methods without tedious recoding or even the knowledge of the bottom level. Finally, the efficiency of the code is also tested through a performance benchmark.}
\end{abstract}

\begin{keyword}
\revised{particle-based \sep general-purpose \sep meshless \sep C++ \sep smoothed particle hydrodynamics \sep discrete element method \sep open source}

\end{keyword}

\end{frontmatter}


{\bf PROGRAM SUMMARY}

\begin{small}
\noindent
{\em Program Title: Nauticle}                                          \\
{\em Licensing provisions: GNU Lesser General Public License v3}                                   \\
{\em Programming language: C++}                                   \\
{\em Nature of problem:}\\
Construction of a flexible simulation tool for particle methods by multilevel user and developer interface for building almost arbitrary mathematical model --- in one, two or three dimensions --- through user-defined algebraic and partial differential equations. \\
{\em Solution method:}\\
At the top level a simulation case can be constructed by \revised{YAML-documents}, hence, the solver does not require any programming knowledge or experience. Besides that, at the second level, the Nauticle environment provides an efficient \revised{flexible} interface in C++ for the adoption of truly arbitrary new schemes. Nauticle can be extended through its C++ interface with any particle method or mathematical model interpretable as a description of relationship between particles by considering them as a set of interaction laws. \revised{The} collection of particle methods and mathematical models \revised{already implemented is}:
\begin{itemize}
  \item Gravitational interaction (for n-body problems)
  \item Smoothed Particle Hydrodynamics (SPH)
  \item Discrete Element Method (DEM)
  \item Discrete Vortex Method (DVM)
  \item \revised{micro-scale social force model}
\end{itemize}
{\em Restrictions:}\\
The present version of Nauticle does not involve implicit schemes.\\
{\em Additional comments:}\\
The Nauticle source code is available at www.bitbucket.org/nauticleproject\\
\end{small}

\section{Introduction} \label{sec:introduction}
Due to their attractive properties, particle-based numerical methods enjoy increasing attention in many fields of engineering applications. In contrast to mesh-based methods like the Finite Element Method (FEM), particle-schemes have more flexible and adaptable spatial discretization of the computational domain of any shape, especially in case of large deformations involving topology changes even with domain splitting \cite{Shaofan2007}.
As far as the implementation is considered, most of the \revised{characteristic} features of particle-based numerical schemes are fundamentally different from \revised{that of} mesh-based methods. Some of these differences are the lack of internodal structure (mesh), the persistent changing of nodal connectivity and the overlapping spatial covering of \revised{the computational domain}. \revised{Relying on such differences, most of the advances related to particle methods show that being robust in different circumstances makes meshless schemes complementary rather than competing with mesh-based methods, and vice versa. As a confirmation, exploiting their aforementioned unique features, particle methods have been applied for the simulation of molecular dynamics \cite{Rapaport2004}, free surface, and multiphase fluid flows \cite{Monaghan1994, Hu2006}, granular materials \cite{Cundall1979} and even fracture of solid structures \cite{Benz1995, Rabczuk2007, Tan2009, Ha2010} with great success, where mesh-based methods suffer from severe and often fateful bottlenecks. Furthermore, multiphysical problems have been investigated as well, incorporating computations and methods for different phenomena \cite{Antoci2007, Ren2013, Robinson2014} and even the coupling with mesh-based Lagrangian and Eulerian methods \cite{Fourey2010, Groenenboom2010, Marrone2016}.}

\revised{Although the concept of particle-based methods is roughly as old as mesh-based models, it was only in the past few decades when they have been advanced significantly \cite{Shaofan2007}. Some of the major problems, such as the intensive computational requirement seem to be gradually solved by the rapidly increasing available computational power of CPU and GPGPU devices \cite{Shaofan2007, Dominguez2011, Govender2016}, others (e.g., particle methods often deal with weaker mathematical background) still need more work to gain upon mesh-based methods. Nevertheless, the present continual development of meshless methods makes them more and more robust and reliable methodologies in several areas of scientific and engineering applications, however, as it is discussed later in this section, the application of the most recent models often requires additional effort from the users.}

\revised{To test and employ the state of the art theories, several tools have been created aiming accuracy and efficiency, though, mostly they are focusing on specific methods and mathematical models in particular. A very few of these open source software packages are DualSPHysics \cite{DualSPHysics}, GPUSPH \cite{GPUSPH} and GADGET \cite{GADGET2} operating with one of the most popular particle-schemes Smoothed Particle Hydrodynamics (SPH), others, such as ESPResSo \cite{ESPResSo}, LAMMPS \cite{LAMMPS}, and GROMACS \cite{GROMACS} provide the fast and reliable solution of molecular dynamics equations on small-scale. Also, open source, three implementation examples for the widely used Discrete Element Method (DEM) are Blaze-DEMGPU \cite{Govender2016}, Yade \cite{Yade} and LIGGGHTS \cite{LIGGGHTS}, from which the latter depends on the high-performance particle engine of LAMMPS. Meeting the requirements on their fields, most of these high-quality tools are more or less being kept up-to-date concerning both the embedded theories and computational practices. Yet, on the one hand, open source simulation tools inevitably reflect the developers' research areas leaving the users alone with their ideas concerning modifications on either of the mathematical models or particle methods. Unfortunately, this approach leads to the grueling process of recoding or further and further implementations of new simulation tools with slightly different and -- at the same time -- usually more focused purposes. On the other hand, extending the capabilities of a code by additional built-in mathematical models appears to be more like a form of further specialization (or at most, a highly fragile generalization) and an endless pursuit, especially in the case of multiphysical capabilities, where not only the governing equations but the numerical methods to be applied may also differ. 

As a novel concept to overcome this issue in scientific computations, general-purpose libraries like FEniCS \cite{fenics} and deal.II \cite{deal2} have been created to facilitate the solution of completely user-defined partial differential equations (PDEs) using mesh-based finite element theories. The basic idea of such libraries is that instead of burning the mathematical models into the source code they are being completely separated or emerged from it, hence users obtain sufficient freedom to configure the desired system of PDEs through a suitable user interface without making serious programming efforts at a lower level. Both FEniCS and deal.II are widely used libraries to investigate a great variety of problems including biomedical and chemical research as well as geosciences \cite{Wilson2014, Hake2012, Schulz2010, Ha2014}.

Similarly to the concepts of FEniCS and deal.II, the same approach has been started to evolve in the field of particle-based computations during the past few years. The first steps in creating such open-source computational environments made by D. Kauzlarich et al. with the code SYMPLER \cite{SYMPLER} and most recently by M. Robinson and M. Bruna with Aboria \cite{Aboria}. Both tools aim generality; SYMPLER can be operated through XML-formatted configuration files, while Aboria is a modern lightweight (header only) library relying on the standard C++ features. They support a collection of particle methods like SPH, DEM and  Molecular Dynamics (MD). Since SYMPLER supports precompilation of user-defined expressions, the computational efficiency is expected to be similar to that of Aboria, however, according to the author's knowledge, neither of their current versions support parallel solutions. Although both solvers provide outstanding freedom compared to the conventional tools from particle methods and governing equations point of view, they suffer from a few drawbacks at the same time. Aboria -- as a library -- supports the most modern practices and user-friendly implementations in C++ and shortens the implementation and application of particle methods dramatically by the built-in domain specific language (DSL), but still requires effort to make in C++ coding and consequently, compilation at every turn. The user interface of SYMPLER helps to circumvent this problem by providing a purely text-based interface for configuration, which is, however, appears to be complicated and requires experience and a relatively good knowledge of the software structure itself.
As a consequence, users need to be familiar with the structure of the code, and a significant amount of non-evident manual work is required to build simulations with both particle simulation engines SYMPLER and Aboria.

The present work proposes Nauticle, a novel, particle-based, open source simulation tool, that attempts to synthesize the most important aspects of generality: the flexible and self-evident configuration of arbitrary mathematical models and the easy implementation of precompiled arbitrary particle interactions separately from the core of the solver. To satisfy these requirements, Nauticle provides three distinct levels for users and developers (from bottom to top):
\begin{itemize}
  \item[] 1. Core developer level: the back end of the tool, including containers, algorithms and fundamental code structure.
  \item[] 2. Intermediate level for developers of particle methods: developer interface for implementation of new particle interactions in C++,
  \item[] 3. Top level for applications (user level): user interface where the mathematical model can be specified using the self-evident Symbolic Form Language (SFL) in a text document based on YAML syntax,
\end{itemize}
The roles and operations of the three levels are introduced in detail in Sections \ref{sec:definitions_and_implementation}, \ref{sec:modeling_workflow} and \ref{sec:implementation_of_interaction_law}. Such a code structure avoids conflicts between developers and users of different levels, which makes Nauticle a flexible and extendable environment for state of the art particle-based methods and completely configurable arbitrary mathematical models (in one, two or three dimensions) through user-defined equations.

The currently implemented numerical methods (particle interactions) in Nauticle are:
\begin{itemize}
  \item Smoothed Particle Hydrodynamics (SPH)
  \item Discrete Element method (DEM)
  \item Gravitational interaction (for N-body simulations)
  \item Discrete Vortex Method (DVM)
  \item micro-scale social force model
\end{itemize}
Note that the inclusion of multipurpose schemes in a single environment significantly facilitates not only the flexible modeling of almost arbitrary problems but the efficient simulation of coupled problems as well.

The paper is organized as follows:
In the second section the main concept of Nauticle with the mathematical background is introduced, then in the subsequent sections \ref{sec:definitions_and_implementation} and \ref{sec:simulation_workflow} the most important definitions are described with the simulation workflow. In Section \ref{sec:modeling_workflow} the usage on the top level is presented through simple examples using the formerly implemented and available particle interactions. Following the top level, the implementation of a new particle scheme is introduced at the intermediate level in Section \ref{sec:implementation_of_interaction_law}. The performance benchmark is presented in Section \ref{sec:performance_benchmark}, then in the remaining two sections \ref{sec:future_improvements} and \ref{sec:conclusions}, the future of the code and further development potentials and the conclusions of the present contribution are discussed.}

\section{Basic idea} \label{sec:basic_idea}
\revised{As it has been partially referred in \cite{SYMPLER} and \cite{Aboria}, particle methods can be interpreted as collections of mathematical expressions describing particle interactions that actually govern the evolution of all particles included in a simulation. Throughout this section, the fundamentals of the general mathematical formulation compatible with Nauticle is introduced.}
\subsection{Definition of the general formulation} \label{sec:formulation}
Consider a set of $N$ spatially distributed point-like objects (hereinafter referred to as particles $p_i$) forming a discrete frame of reference called particle system $P_N$ inside a one, two or three dimensional axis-aligned rectangular domain $D$. Assign arbitrary fields of zeroth, first or second order tensorial quantities $\Phi^\alpha$ to $P_N$:
\revised{
\begin{equation} \label{eq:field}
\begin{split}
&\Phi^\alpha:P_N\rightarrow {\rm I\!R}^{d_1\times d_2} \quad \text{with } d_1,d_2=1,2,3 \\
&\Phi^\alpha_i=\big\{\Phi^\alpha(p_i)\big\vert p_i \in P_N\big\},
\end{split}
\end{equation}
where greek letters ($\alpha, \beta ...$) are used to index different fields. Obviously, the positions of the particles can also be expressed as a tensorial field.
Let us assume, that $\Phi^\beta$ depends on 
\begin{equation}
\Phi^I:P_N\rightarrow {\rm I\!R}^{d_1\times d_2} \quad (I=1,2,...,M)
\end{equation}
fields and
\begin{equation}
\Psi^a:P_N\times P_N\rightarrow {\rm I\!R}^{d_1\times d_2} \quad (a=1,2,...,K)
\end{equation}
interaction laws such that
\begin{equation} \label{eq:general_interaction}
\Phi^\beta_i=\frac{d\Phi^{\alpha}_i}{dt}=f\big(\Psi^1_i,\Psi^2_i,...,\Psi^K_i\big)+g\big(\Phi_i^{1},\Phi_i^{2},...,\Phi_i^M\big),
\end{equation}
with $f,g:{\rm I\!R}^{d_1\times d_2}\rightarrow{\rm I\!R}^{d_1\times d_2}$, while $\Psi^a_i$ has the form of:
\begin{equation} \label{eq:interaction_formula}
\Psi^a_i=\sum_{j=1}^{n(\Delta_a)}{I_a\big(\Phi_i^{1},\Phi_i^{2},...,\Phi_i^{M},\Phi_j^{1},\Phi_j^{2},...,\Phi_j^{M}\big)},
\end{equation}}
where $n$ is the number of neighbors around particle $i$ depending on the finite or infinite influence radius $\Delta_a$, and $I_a\big(\Phi_i^{1},\Phi_i^{2},...,\Phi_i^{M},\Phi_j^{1},\Phi_j^{2},...,\Phi_j^{M}\big)$ is considered to be an appropriate interaction law (interaction operator) between particles $i$ and $j$. Thus, the first term $f\big(\Psi^1_i,\Psi^2_i,...,\Psi^K_i\big)$ on the right hand side implies pair-interactions depending on the spatial configuration of the particles, while the second term $g\big(\Phi_i^{1},\Phi_i^{2},...,\Phi_i^M\big)$ represents particlewise expressions omitting neighboring particles. The construction of \equref{eq:general_interaction} for each particle leads to a set of $N$ ordinary differential equations (ODEs).

Nauticle is designed to solve such problems that are directly governed by a system of the form of the ODE in \equref{eq:general_interaction} or can be transformed (through a suitable discretization scheme) to the same form. In other words, for instance, the governing equations obtained with the famous meshless collocation technique SPH \cite{Gingold1977,Monaghan1992,Gray2001,Bui2008} or the collision modeling DEM \cite{Cundall1971,Taylor2006} and many other particle-based schemes (even the stochastic Dissipative Particle Dynamics (DPD) scheme \cite{Hoogerbrugge1992}) meet the ODE in \equref{eq:general_interaction} offering the potential to construct and solve them similarly through numerical integration.

\revised{Using the top level of Nauticle, the complete configuration of the simulated mathematical model as user-defined equations is straightforward -- by the definition of symbols $\Phi^\alpha$ and the application of precompiled interaction operators $I_a\big(\Phi_i^{1},\Phi_i^{2},...,\Phi_i^{M},\Phi_j^{1},\Phi_j^{2},...,\Phi_j^{M}\big)$ of equation \equref{eq:general_interaction} -- through a purely text-based configuration file without any programming and recompilation. Furthermore, the intermediate developer level makes the implementation of precompiled interaction operators separated from the core of the code (cf. Section \ref{sec:implementation_of_interaction_law}).}

\section{\revised{Definitions and implementation}} \label{sec:definitions_and_implementation}
\revised{Throughout this section, the definitions and implementations facilitating the solution of equation \equref{eq:general_interaction} are presented. The quantities in \equref{eq:general_interaction}, the real-valued scalar, vector or tensorial symbols $\Phi^\alpha$ are managed by the unified double precision tensor-container, hence, unless otherwise indicated, values of symbols and expressions are hereinafter considered to be instantiations of the unified tensorial quantity.}

\subsection{\revised{Symbolic Form Language (SFL), class hierarchy of expressions}} \label{sec:class_hierarchy_of_expressions}
\revised{The interpretation of the user-defined expressions written using SFL in the configuration file is based on the class hierarchy of the expressions shown in Figure \ref{fig:class_hierarchy} (SFL-tree). White nodes of the diagram denote abstract types. Obviously, the two main branches are initiated by the abstract Symbol and Operator classes, from which further types are inherited.

Nauticle provides different types of user-defined symbols. Therefore, the symbol branch consists of the container classes such as the single-valued Constant and Variable, as well as the Field and Particle system classes, which are, on the contrary, allowed to store values separately for each particle. Besides storing the particle coordinates, the Particle system class is also responsible for the neighbor search.

Nodes of the operator branch are also divided into different branches: arithmetic functions and operators are performing particle-wise computations disregarding the particle layout, while the operators depending on the position and neighborhood data stored in the particle system are inherited from the also abstract Interaction class. Avoiding the visualization of too many nodes with similar roles, blue boxes in the hierarchy denote numerous particle-wise operator nodes wrapped together. The dependency of the interactions on the particle data is expressed in the diagram by the dashed line. Among other operators such as search of extrema, the implementations of particle methods are also derived from the Interaction class. Thus, as it is discussed later in Section \ref{sec:implementation_of_interaction_law}, the Interaction node is also considered to be the interface of Nauticle for developers on the second, intermediate level.
\customfigure{0.5}{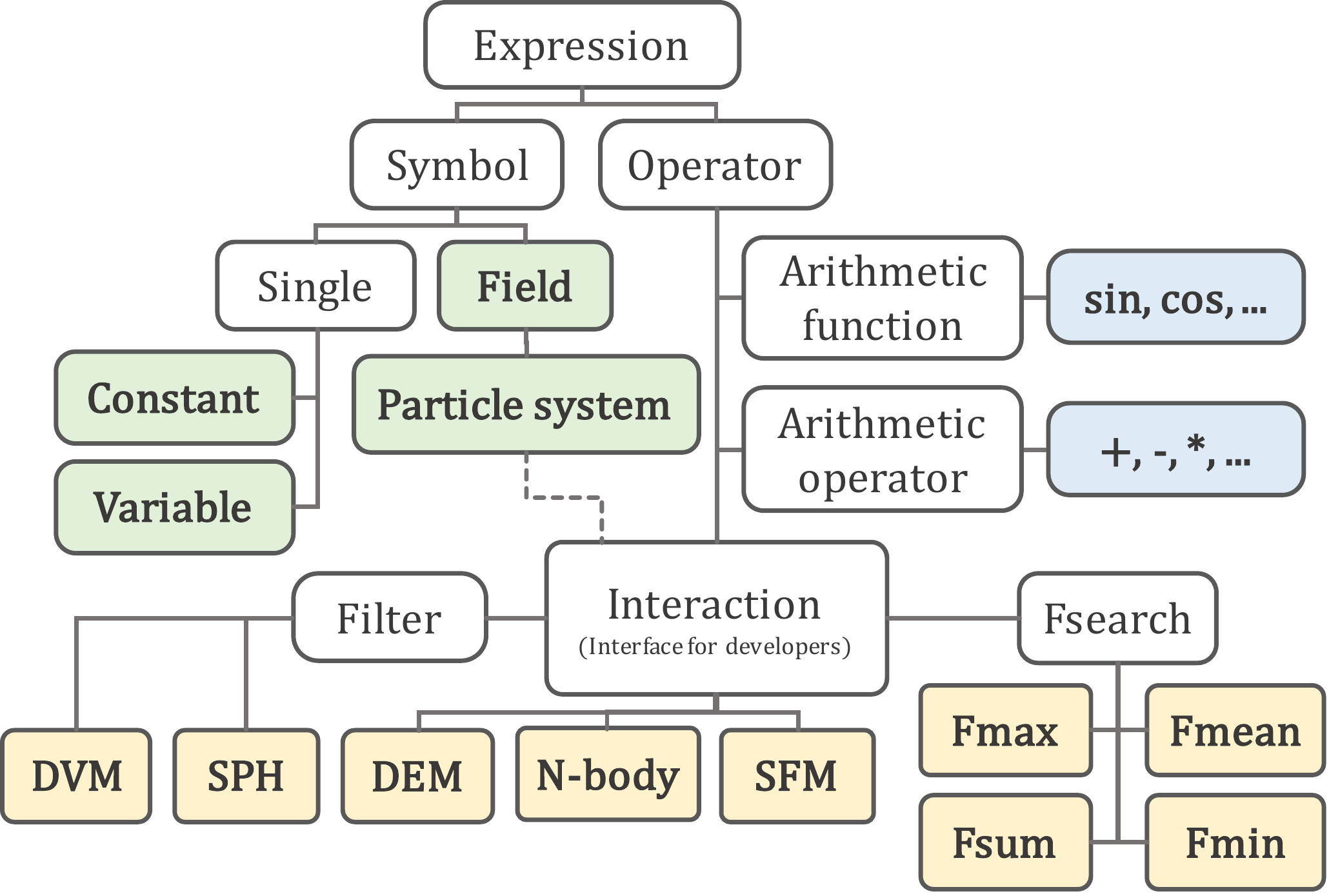}{\revised{Class hierarchy applied to parse user-defined equations. White nodes represent abstract types.}}

During the execution, using the class hierarchy, user-defined expressions are built and recursively evaluated at runtime as it is demonstrated in Figure \ref{fig:recursive_execution}.
\customfigure{0.5}{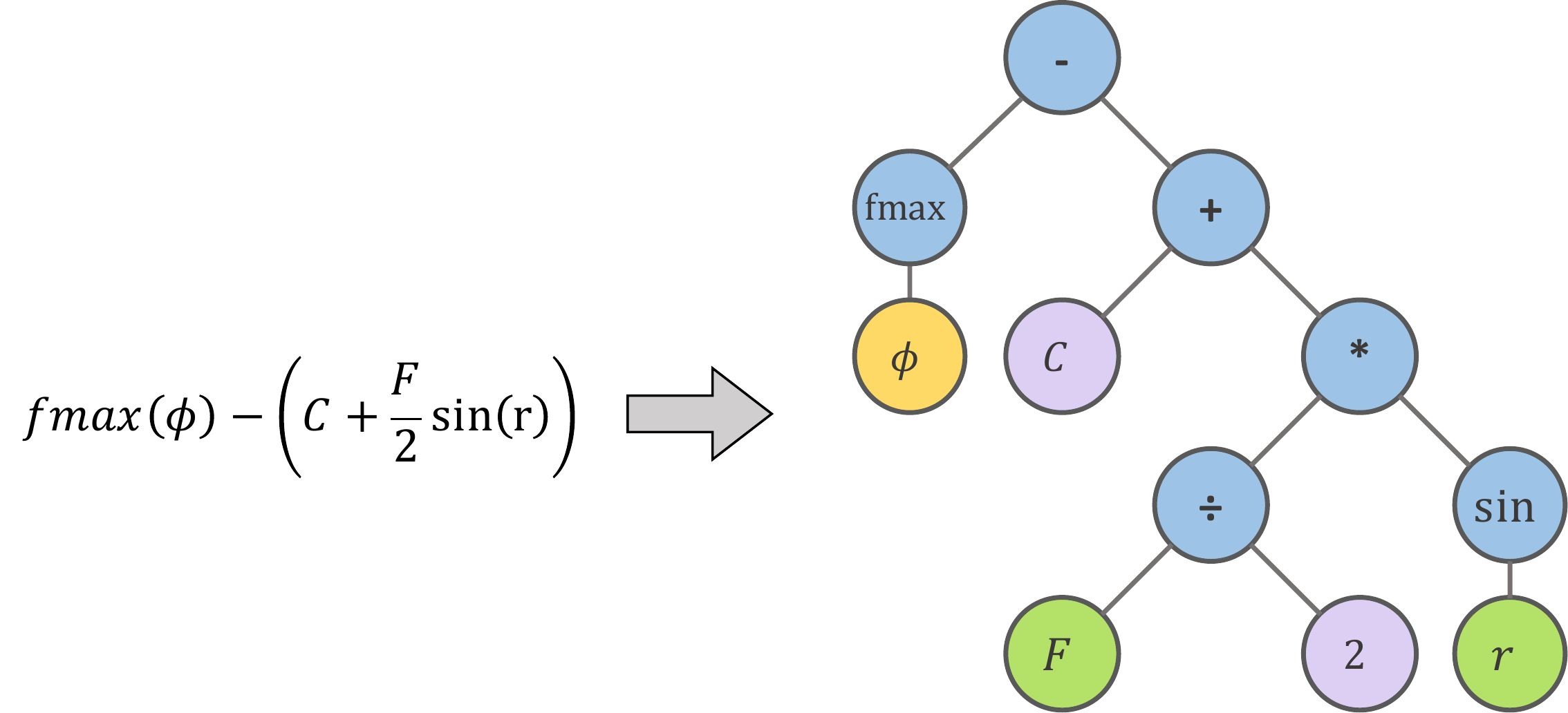}{\revised{Example of expression tree built from user-defined expression. Blue, green, yellow and purple nodes represent operators, variables, fields, and constants respectively.}}

Since the field and the particle system nodes are containers of multiple values -- assigned to particles, the evaluation of expressions including symbols of field or particle system is also eventuated in a field quantity. Evaluation of expressions consisting of different symbols such as fields and variables is also possible.

In contrast with SYMPLER \cite{SYMPLER} and Aboria \cite{Aboria}, due to the hardcoded particle interaction laws of the SFL-tree in Figure \ref{fig:class_hierarchy}, the desired numerical model can be constructed significantly more intuitively, while maintaining sufficient flexibility (for details, see the examples in Section \ref{sec:modeling_workflow}).

\subsection{Simulation case} \label{sec:simulation_case}
As an explanation of the solution process, Figure \ref{fig:implementation} shows a schematic diagram of the interpretation and storage of user-defined symbols and equations, as well as the functions solving them. Firstly, as it has been explained in the previous subsection, the equations and symbols given as strings are interpreted by the SFL expression parser. After the interpretation, the symbols are getting stored in the vector container inside the so-called workspace object. As the owner of all quantities describing the mathematical model, the workspace is responsible for the management of symbols with methods such as sorting, printing and value querying of any symbol. Together with the workspace, equations are also stored in the object called Case. Each of the equations consists of two objects: a left-hand side (LHS) and a right-hand side (RHS), representing a symbol and an expression respectively. The solution process of any equation performs the evaluation of the RHS, from which the result then overrides the data stored in the LHS symbol. Due to the shared owner semantics applied between the LHS and the workspace, the symbols are immediately updated in the workspace without any further user actions. It is worth to mention that it is not monitored at any time if the equations are given in implicit form, hence, in such a case when an interaction is dependent on the LHS symbol, the solution becomes unpredictable. Race condition issues are also discussed in Sections \ref{sec:multithreading} and \ref{sec:implementation_of_interaction_law}.

As it is also shown in the diagram, the solution of the equations is driven by the case object by iterating through the vector of equations and executing their solve functions one by one.

\customfigure{0.5}{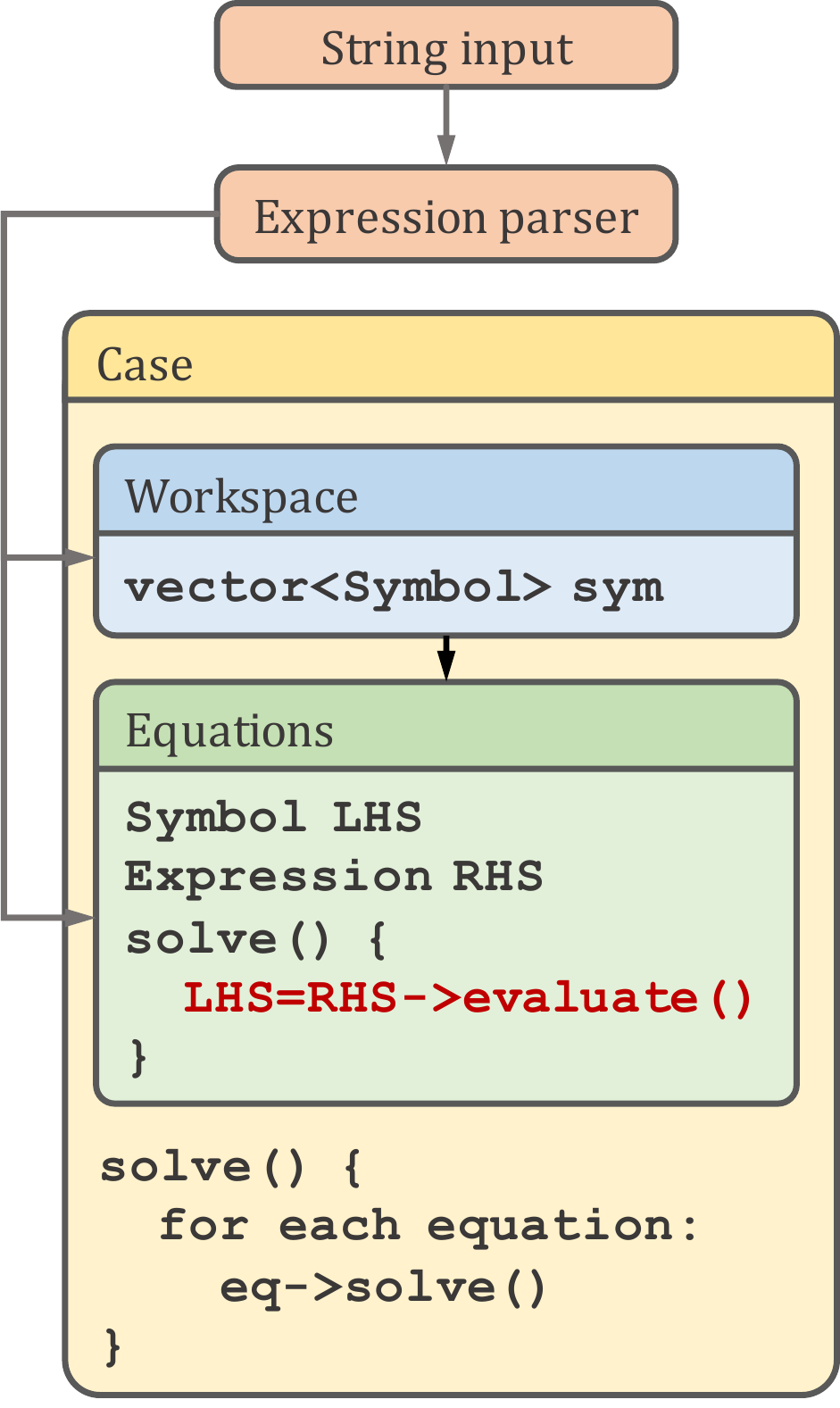}{\revised{Interpretation and storage of user-defined symbols and equations in the Case object. The solution of the equations is performed in the line highlighted by red.}}}

\subsubsection{The computational domain} \label{sec:domain}
The computational domain of any particle simulation in Nauticle is considered to be a \revised{one-, two-, or three-dimensional} dimensional axis-aligned box, in which the particle system $P_N$ is interpreted, hence the neighbor search is performed. By definition, the existence of particles is not allowed out of the domain's volume. To support the neighbor search algorithm, the domain is divided into \revised{cuboid} cells of user-defined edge \revised{sizes $\rm{\Delta}_e$} (practically greater or equal than the influence radii of the particles) covering the computational volume with a \revised{rectilinear} spatial grid. Since due to the cell-based neighbor search, pairs with interparticle distance larger than \revised{$\rm{\Delta}_e$} \revised{in any directions} are consistently ignored, \revised{the cell size} is recommended to be equal to the influence radius of the applied interaction law \revised{($\Delta_a$)}. The domain size and grid layout \revised{are} defined at the start and remain unchanged during the whole simulation.

Regarding the particle interactions and motions close to the domain surfaces, three different types of boundary treatments are possible: \textit{periodic}, \textit{symmetric} and \textit{cut-off} \revised{boundaries}. Obviously, the opposing bounding surfaces of the domain need to possess identical boundary conditions. Additionally, particles crossing any of the bounding surfaces -- hence leaving the computational domain -- are shifted periodically, however, in case of symmetric boundaries it should never occur. As the simplest model, cut-off surfaces omit any specific particle treatment except for the periodic particle shifting.

\subsection{\revised{Multithreading}} \label{sec:multithreading}
\revised{Due to their specialized programming requirements, it is apparent, that the solution of the user-defined equations in interpreter mode (even if the interactions are already compiled) obstructs the efficient utilization of massively parallel devices such as GPGPUs. However, widely spread multicore CPU devices are also potentially applicable for parallelization of particle-computations in a less challenging manner \cite{Napoli2015,DualSPHysics}.

Since Nauticle is written in C++11, the built-in standard multithreading library is a reasonable choice for the parallel evaluation of expressions even in interpreter mode. Unlike the parallelization based on the decomposition of the domain in \cite{Napoli2015}, Nauticle is designed to solve expressions with the decomposition of the particle system in such a way that the number of particles per thread $n_{ppt}$ is calculated as
\begin{equation}
  n_{ppt}=\frac{N_p}{N_t},
\end{equation}
where $N_p$ is the total number of particles and $N_t$ is the number of threads (equal to the number of hardware-supported threads by default). To avoid race conditions by restricting the number of simultaneously modifiable symbols to one, no solution of any of the equations is started until the previous equation has been completely finished for all particles. Furthermore, none of the particles are treated by more than one thread at the same time. Obviously, in the special case when the LHS covers a variable, only a single thread is used for evaluation.}

\section{\revised{Solution process}} \label{sec:simulation_workflow}
\revised{In this section, the simulation process involving the case structure (cf. Section \ref{sec:simulation_case}) is presented from a higher perspective. The schematic diagram and flow of the solution in Nauticle is explained by Figure \ref{fig:simworkflow}.

Every simulation case in Nauticle is defined by a single YAML file; however, in case of hot-start simulations, a previous result file is also required as an initial condition (left-hand side region of Figure \ref{fig:simworkflow}). Furthermore, particle positions are allowed to be imported from an external source using an optional ASCII data storage file. Optional files of both the nodal positions and initial conditions are accessed by their names defined in the YAML file. As it has been already shown in \ref{sec:simulation_case}, before the assembly of the case, the SFL expression parser translates the extracted data of the workspace and the user-defined equations and passes them to the case assembler. Once the case (illustrated by the orange box on the right-hand side) has been successfully built, the solution process presented in Section \ref{sec:simulation_case} is managed by the simulation scheduler based on the simulation parameters also defined in the YAML file. The simulation case object including both the workspace and the list of equations is written to ASCII or binary formatted VTK result files with a pre-defined frequency through the VTK writer. Being too specific data structures for VTK format, some parts of the simulation case (the computational domain, single symbols of the workspace and the user-defined equations) are converted to character strings and automatically retrieved as initial conditions in further simulations.
\customfigure{0.4}{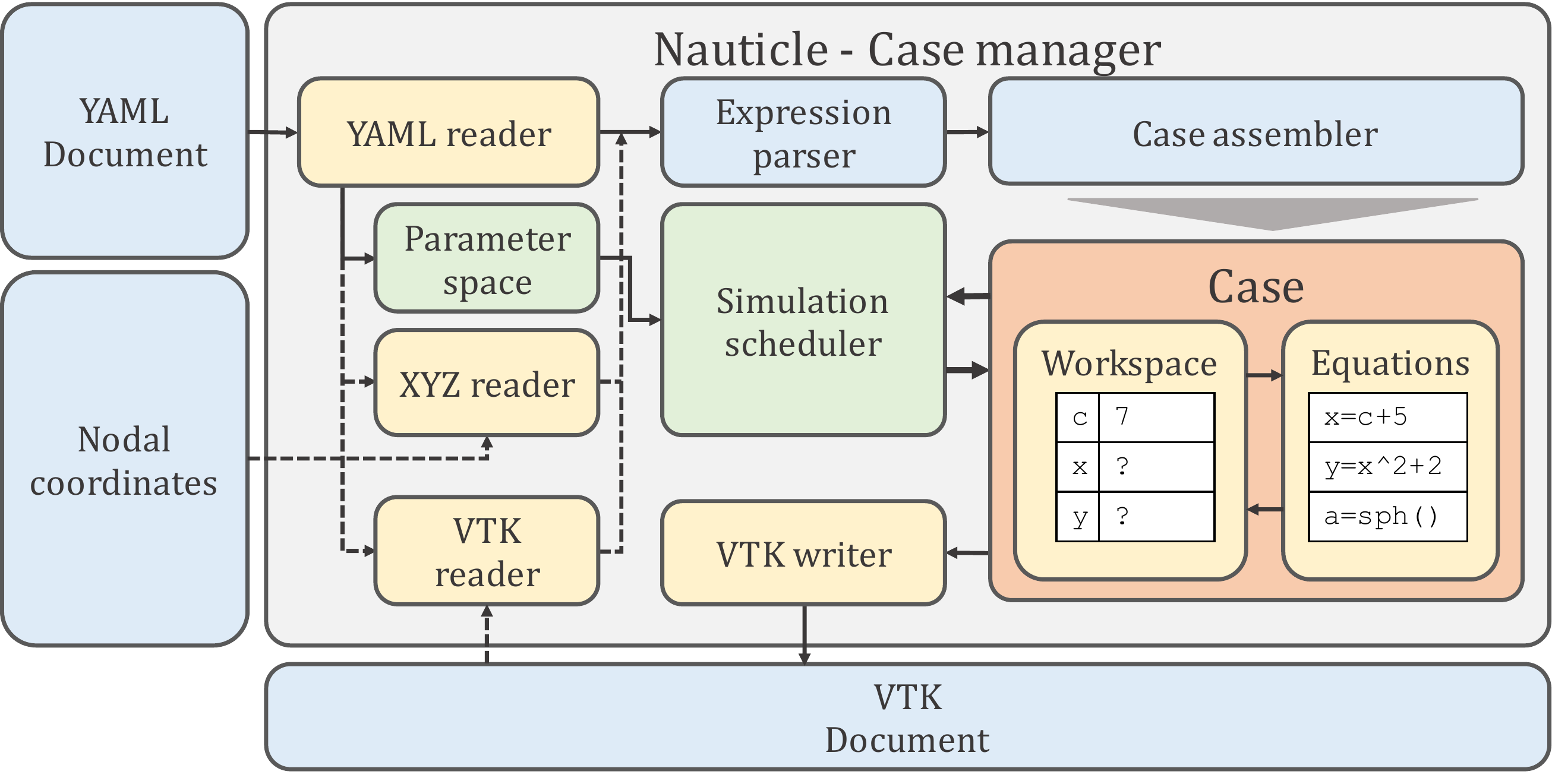}{Schematic diagram of the Nauticle simulation flow.}

One important part of the simulation is the proper execution of the neighbor searching methods. Since the equations are defined by the user, it is not known a priori, when to update the list of neighbors. Furthermore, in case of constant particle positions the neighbor search needs to be run only once in the beginning of the simulation. Hence, to avoid unnecessary waste of computational time, but at the same time, perform the searching process once it is required, the neighbor search methods are linked to the simulation scheduler and executed periodically only if the particle layout has been changed.}

\section{Modeling workflow and examples} \label{sec:modeling_workflow}
\revised{The workflow presented in this section is restricted to the top level of Nauticle, assuming that the particle schemes in question are already implemented in the SFL hierarchy of expressions. Such cases, when the interaction laws of a numerical model are not yet adopted in Nauticle are discussed in Section \ref{sec:implementation_of_interaction_law}.}

To configure and run a calculation using Nauticle, the following steps should be performed in order:
\begin{itemize}
  \item Construct governing equations to the desired problem and list the symbols required to describe the \revised{mathematical} model.
  \item Choose suitable numerical schemes for the given equations.
  \item If not yet adopted, implement a class of interaction laws for the desired numerical scheme in C++ (cf. Section \ref{sec:implementation_of_interaction_law}) and connect it to Nauticle \revised{through the Interaction node of the SFL-tree}.
  \item \revised{Write the symbols and equations using the interaction laws (e.g., by discretization) in SFL-format.}
  \item Construct the configuration YAML file with the definition of the case including the workspace and equations.
  \item Set parameters for the simulation and output data.
  \item Run Nauticle to perform the calculation with the pre-defined \revised{YAML-document}.
\end{itemize}

\revised{The following Section \ref{sec:example1a} guides the reader through the steps above by building a two-dimensional SPH dam break simulation, while the Section \ref{sec:example1b} explains only the required changes of example $1a$ to construct a fundamentally different simulation from both geometrical and mathematical model point of views.}

\subsection{\revised{Example 1 (SPH)}}
\revised{The particle method applied in both examples $1a$ and $1b$ is chosen to be SPH. The method has been firstly published in 1977 by R.A. Gingold and J.J. Monaghan \cite{Gingold1977} and independently by L.B. Lucy \cite{Lucy1977}, however, the most significant advances in the topic were made in the past twenty years mostly in modeling fluid flows and astrophysical problems \cite{Monaghan2005}. Using SPH, similarly to the Finite Difference Method (FDM), the derivatives and consequently the governing equations are approximated using algebraic expressions, however, without the strict and rigid underlying grid of sample points.}

\revised{The basic idea of the approximation is deduced from the generalized interpolation
\begin{flalign} \label{convolution_delta}
  A(\textbf{r})=\int_{\Omega}{A(\textbf{r}')\delta(\textbf{r}-\textbf{r}')dV},
\end{flalign}
where $dV$ is the differential volume element, $A$ is an arbitrary function of the position $\textbf{r}$ over the domain $\Omega$ and $\delta$ is Dirac's function often substituted with a numerically feasible mollifier or smoothing kernel function $W(\textbf{r}-\textbf{r}',\Delta)$ of finite or infinite influence radius $\Delta$ \cite{Monaghan2005}. Furthermore, the continuum integration can be discretized to obtain the fundamental SPH sampling expression over a set of point-like discrete nodes
\begin{flalign} \label{convolution_delta}
\langle A(\textbf{r}_i)\rangle=\sum_j{A(\textbf{r}_j)W(\textbf{r}_i-\textbf{r}_j,\Delta)\frac{m_j}{\rho_j}}=\sum_j{A_jW_{ij}V_j}.
\end{flalign}
Without the detailed representation provided by \cite{Violeau2012}, the first and second order spatial derivatives of $A$ are expressed respectively as
\begin{flalign} \label{eq:diffop1}
  \langle \nabla A_i\rangle=\rho_i\sum_{j}{\bigg(\frac{A_i}{\rho_i^2}+\frac{A_j}{\rho_j^2}\bigg)m_j\nabla W_{ij}},\\\label{eq:diffop2}
  \langle \nabla A_i\rangle=\sum_{j}{(A_j-A_i)\frac{m_j}{\rho_j}\nabla W_{ij}},\\ \label{eq:diffop3}
  \langle \Delta A_i\rangle=\sum_{j}{2\big(A_i-A_j\big)\frac{\textbf{r}_j-\textbf{r}_i}{\vert \textbf{r}_j-\textbf{r}_i \vert^2}\nabla W_{ij}},
\end{flalign}
where $\nabla W_{ij}$ is the analytical derivative of the smoothing kernel function. The \equref{eq:diffop1} and \equref{eq:diffop2} are differential operators satisfying different conservativity and consistency properties for different purposes.}

\subsubsection{\revised{Example 1a - Simulation of a 2D dam break flow}} \label{sec:example1a}
\revised{As a minimal example, one of the most frequently investigated SPH test case, a two-dimensional dam break problem is constructed with symmetric free-slip boundary conditions. The initial layout of the system is shown in Figure \ref{fig:example1a_geom}.
\customfigure{0.7}{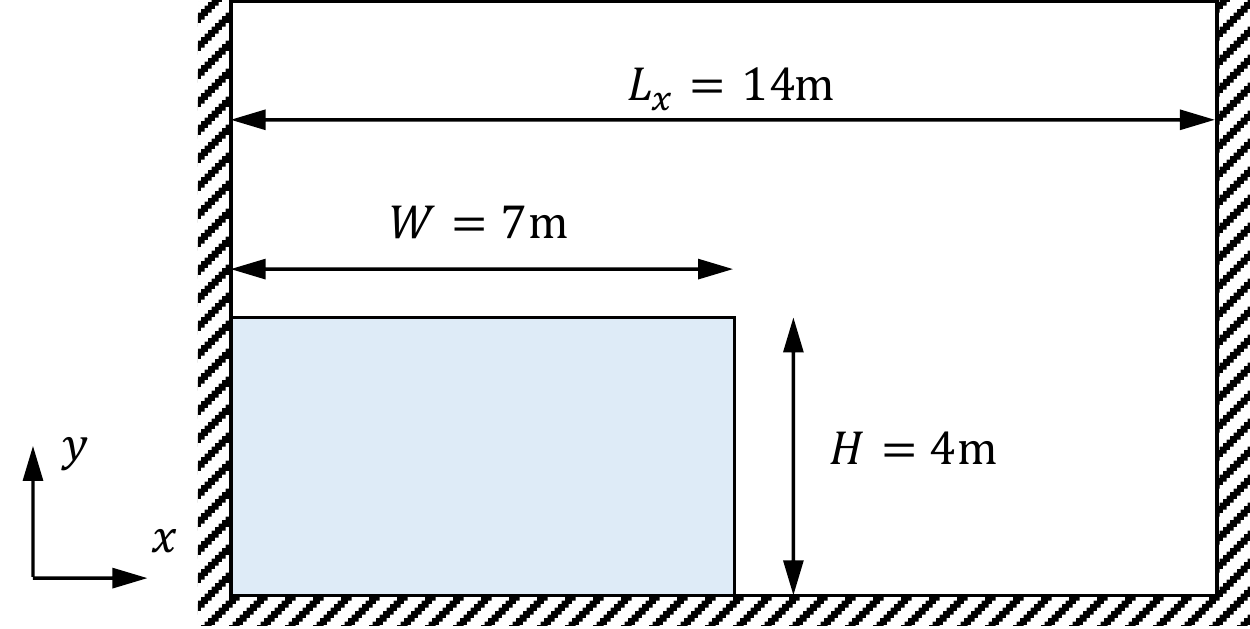}{\revised{Initial configuration of the two-dimensional dam break (\textit{example 1a}).}}
The governing equations of the inviscid weakly compressible free surface dam break flow can be written in Lagrangian frame as
\begin{flalign} \label{eq:fluid_eqs}
\frac{d\rho}{dt}&=-\rho\nabla \textbf{v}, \\
\frac{d\textbf{v}}{dt}&=-\frac{1}{\rho}\nabla p+\textbf{g},
\end{flalign}
where $\rho$, $p$, $\textbf{v}$ and $\textbf{g}$ are the fluid density, fluid pressure, velocity and gravitational acceleration respectively. After the discretization of the equations according to \equref{eq:diffop1} and \equref{eq:diffop2}, the following numerical model
\begin{flalign} \label{eq:SPH_fluid_continuity}
\frac{d\rho_i}{dt}&=-\rho_i\sum_j{(\textbf{v}_j-\textbf{v}_i)\frac{m_j}{\rho_j}\nabla W_{ij}}, \\
p_i&=c^2(\rho-\rho_0), \\
\frac{d\textbf{v}_i}{dt}&=-\sum_j{\bigg(\frac{p_i}{\rho_i^2}+\frac{p_j}{\rho_j^2}+\Pi_{ij}\bigg)m_j\nabla W_{ij}}+\textbf{g},
\label{eq:SPH_fluid_momentum}
\end{flalign}
is obtained, introducing the artificially reduced sound speed $c=50$m/s, the reference density $\rho_0=1000$kg/m$^2$, the constant particle mass $m$ depending on the interparticle distance $dx$, and the artificial viscosity term
\begin{equation}
    \Pi_{ij}= 
\begin{cases}
  -0.1ch\frac{\textbf{v}_{ji}\textbf{r}_{ji}}{\rho_i\vert\textbf{r}_{ji}\vert^2} & \text{if } \textbf{v}_{ji}\textbf{r}_{ji} < 0, \\
  0 &\text{ otherwise},
\end{cases}
\end{equation}
with $\textbf{v}_{ji}=\textbf{v}_{j}-\textbf{v}_{i}$ and $h=1.1dx$ being the smoothing radius of the kernel function $W_{ij}=W(\vert\textbf{r}_{i}-\textbf{r}_{j}\vert,h)$. Here, the widely used fifth order Wendland kernel function is used:
\begin{equation}
W(\vert\textbf{r}_{i}-\textbf{r}_{j}\vert,h)=\alpha_D\bigg(1-\frac{q}{2}\bigg)^4(2q+1),
\end{equation}
where $D$ is the number of dimensions for normalization, $\alpha_1=3/(4h)$, $\alpha_2=7/(4\pi h^2)$ and $\alpha_3=21/(16\pi h^3)$ and $q=\vert\textbf{r}_{i}-\textbf{r}_{j}\vert/h$. Being compatible with the general form \equref{eq:general_interaction}, in the following, the construction of the equations (\ref{eq:SPH_fluid_continuity} - \ref{eq:SPH_fluid_momentum}) with SFL is presented.}

\revised{Firstly, the constant and variable symbols involved in (\ref{eq:SPH_fluid_continuity} - \ref{eq:SPH_fluid_momentum}) need to be defined:}
\begin{lstlisting}[caption={\revised{Definition of constants and a variable}}]
constants:
  - rho0: 1000        # rest density
  - h: 0.25           # smoothing radius
  - dx: h/1.1         # particle spacing
  - c: 50             # speed of sound
  - mass: dx^2*rho0   # particle mass
  - g: 0|-9.81        # gravity
variables:
  - dt: 1e-3          # time step size
\end{lstlisting}
\revised{The separator in \texttt{0|-9.81} indicates column vector with $x$ and $y$ components respectively. The definition and value assignment is carried out in the given order, thereby, facilitating flexibility, SFL supports expressions for definitions depending on formerly defined symbols (e.g. "\texttt{mass: dx\string^2*rho0}"). Also, note that the time step size \texttt{dt} has to be defined in all simulations in Nauticle and it is strongly recommended to be a variable even if it is required to be constant because the simulation scheduler may reduce it temporarily to generate results in predefined instants of the simulation. Obviously, \texttt{dt} is never increased to satisfy such a condition.}

\revised{After the definition of the constants and variables, the particle system can be generated including the domain with cell sizes, and boundary conditions. Since the neighbour search depends on the \texttt{cell\_size}, it is computationally desirable to choose it to be equal to the particle influence radii, which is $2h$ in case of the Wendland kernel function. The \texttt{minimum} and \texttt{maximum} values are given as the integer number of cells in each coordinate direction.}
\begin{lstlisting}
domain:
  cell_size: 2*h|2*h
  minimum: 0|0
  maximum: 7/h|10/h
  boundary: symmetric|symmetric
\end{lstlisting}
\revised{Besides the \texttt{domain}, the initial particle layout should also be given in the particle system using either external text file with particle positions (cf. Section \ref{sec:example1b}) or an axis-aligned rectangular grid with position \texttt{gpos}, size \texttt{gsize} and grid spacing \texttt{gip\_dist} properties and a grid-identifier \texttt{gid} value.}
\begin{lstlisting}
grid:
  gid: 0
  gpos: 0|0
  gsize: 7|4
  goffset: 0|0
  gip_dist: dx|dx
\end{lstlisting}
\revised{The workspace becomes fully defined by the required field quantities, that are also initialized with the given values:}
\begin{lstlisting}
fields:
  - rho: rho0   # fluid density
  - rhodot: 0   # rate of change of density
  - v: 0|0      # velocity
  - vdot: 0|0   # acceleration
  - p: 0        # pressure
\end{lstlisting}
\revised{Additionally, the field \texttt{r} storing particle positions is automatically generated according to the particle system data.}

\revised{In order to complete the configuration, the user-defined equations need to be constructed using the symbols above. The construction process of user-defined equations through SFL is presented in Figure \ref{fig:user_defined_equation_intro}. 
\customfigure{0.5}{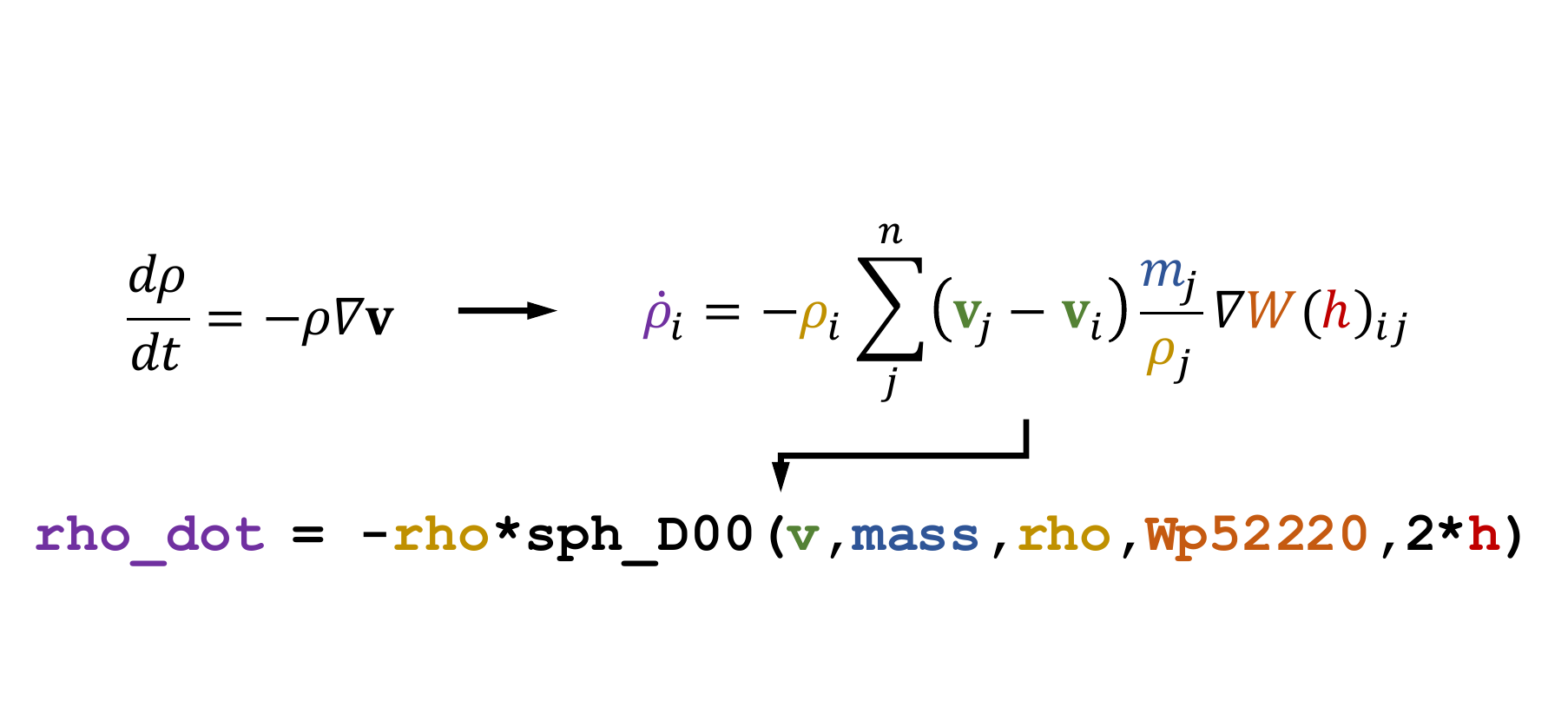}{\revised{Construction of the continuity equation using SFL.}}
In this particular case, the \texttt{sph\_D00} operator name tells the SFL, which interaction law needs to be applied for building the expression tree according to Figure \ref{fig:recursive_execution}. The complete list of the currently implemented interaction operators is available in the user's guide of Nauticle. Similarly, the \texttt{Wp52220} keyword indicates the desired kernel function based on the notation in Figure \ref{fig:kernel_explanation}
\customfigure{0.3}{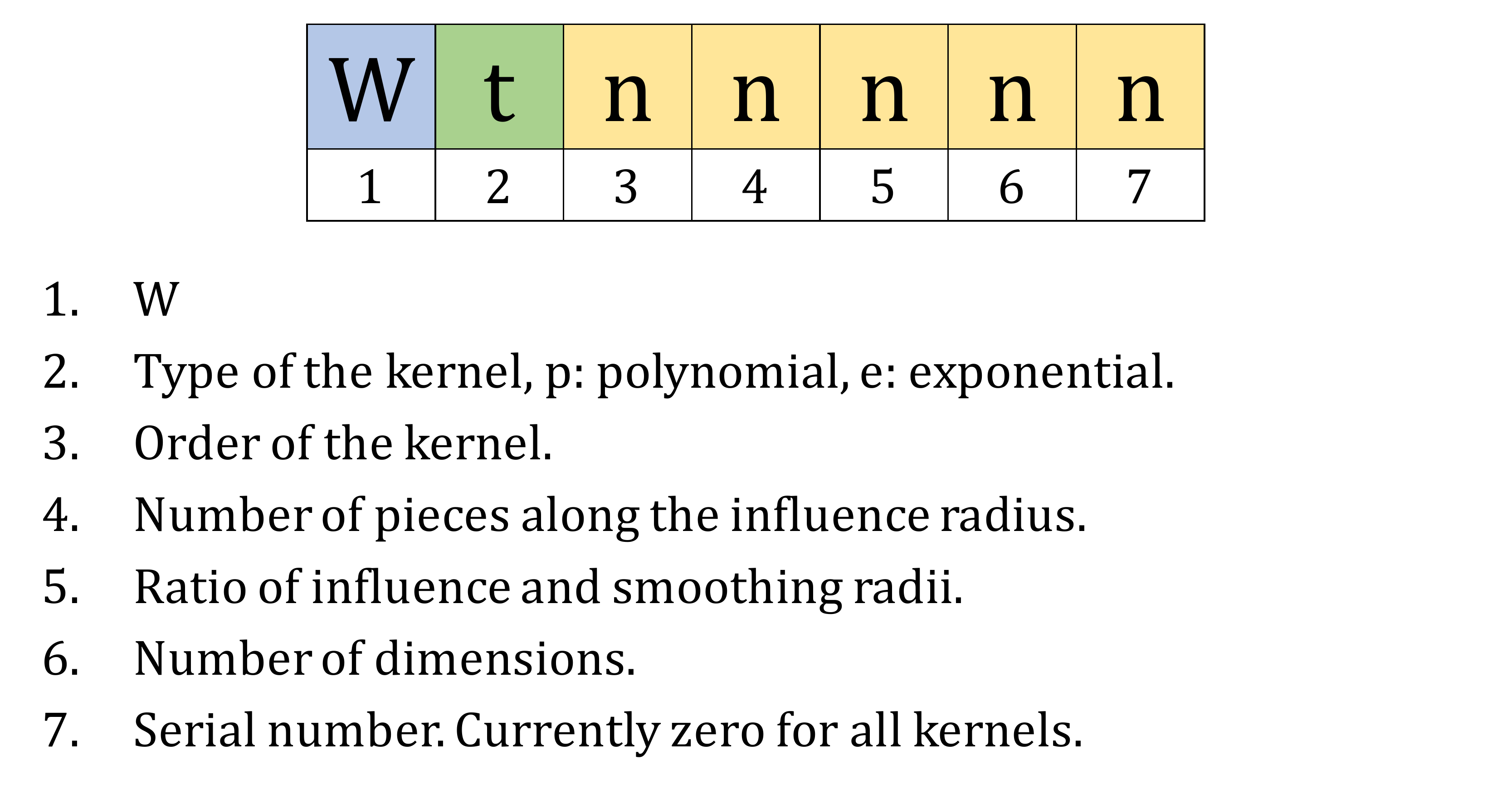}{\revised{Explanation of the keyword to the smoothing kernel function.}}
Thus, the user defined equations to represent the system (\ref{eq:SPH_fluid_continuity} - \ref{eq:SPH_fluid_momentum}) in SFL can be written as}
\begin{lstlisting}
equations:
  - eq1: rhodot=-rho*sph_D00(v,mass,rho,Wp52220,2*h)
  - eq2: rho=euler(rho,rhodot,dt)
  - eq3: p=c^2*(rho-rho0)
  - eq4: vdot=-1/rho*sph_G11(p,mass,rho,Wp52220,2*h)+g
  - eq5: vdot=vdot+0.1*c*h*sph_A(v,mass,rho,Wp52220,2*h)
  - eq6: v=euler(v,vdot,dt)
  - eq7: r=euler(r,v,dt)
\end{lstlisting}
\revised{where equations \texttt{eq2}, \texttt{eq6} and \texttt{eq7} are responsible for the temporal integration of the system, currently using the simplest, first order semi-implicit Euler integration scheme. Note that \texttt{eq4} and \texttt{eq5} are corresponding to the same equation and separated only for the sake of better readibility. Operators \texttt{sph\_G11} and \texttt{sph\_A} refer to the pressure gradient and artificial viscosity respectively in \equref{eq:SPH_fluid_momentum}.
Finally, the simulation time and the interval for printing results are configured in the \texttt{parameter\_space} block:}
\begin{lstlisting}
parameter_space:
  simulated_time: 6
  print_interval: 0.05
\end{lstlisting}
\revised{The complete configuration file of this example is attached in Appendix \ref{app:config_example1a}.

The simulation can be performed by passing the configuration file to Nauticle in command prompt by executing the command \texttt{\$ nauticle -yamlname <configuration file>}. After running the simulation, the vtk result files can be visualized in Paraview (Figure \ref{fig:example1a_result}).}

\customfigure{0.5}{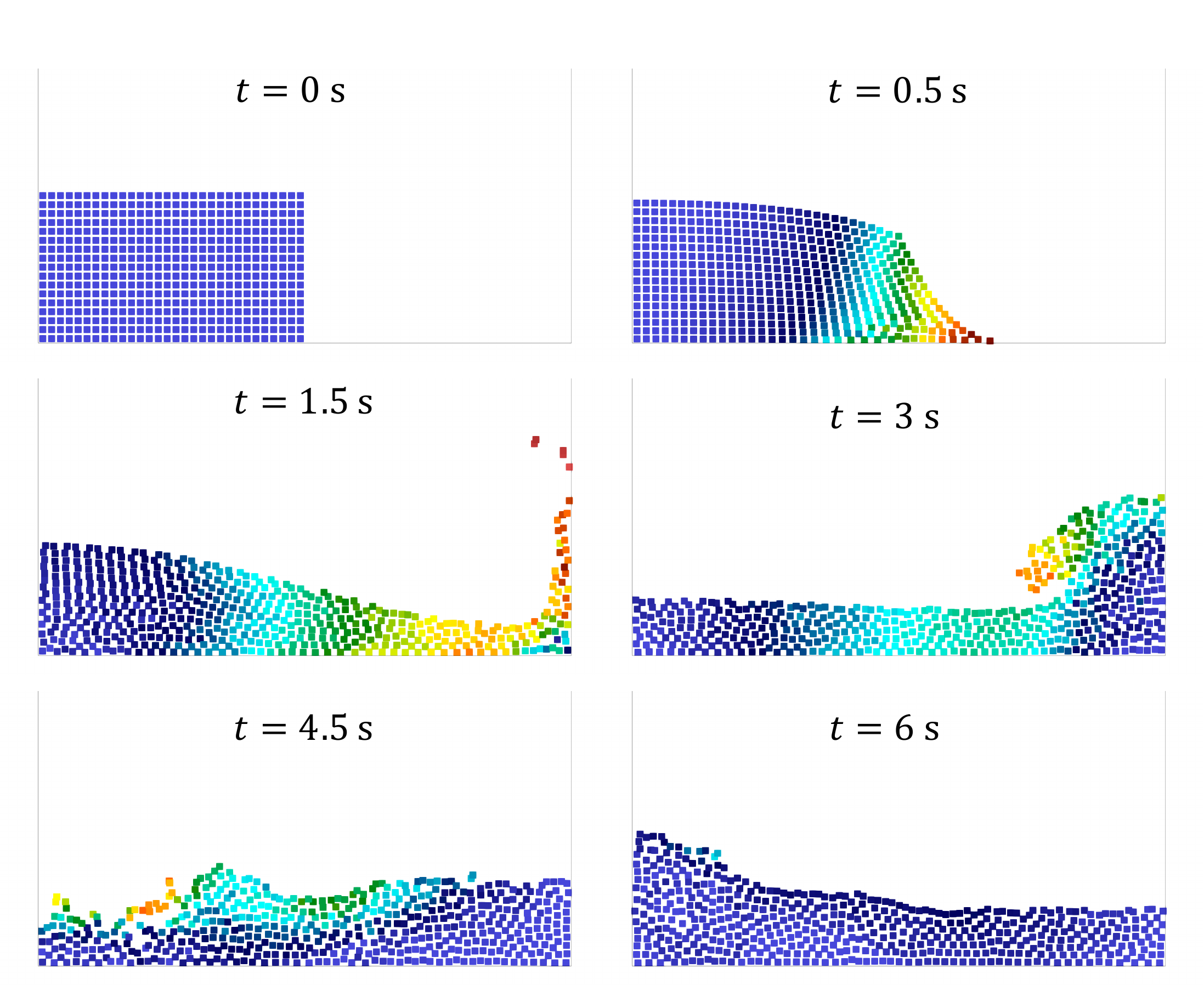}{\revised{Result of the 2D dam break simulation displayed in different time instants.}}

\subsubsection{\revised{Example 1b - Simulation of phase separation on a sphere}} \label{sec:example1b}
\revised{The following SPH example illustrates the flexibility of Nauticle by presenting the most important modification steps of the SFL configuration of \textit{example 1a} to obtain the fundamentally different numerical model.

The process of the separation of two phases is usually modelled by the fourth order nonlinear Cahn-Hilliard equation
\begin{flalign} \label{eq:cahnhilliard}
\begin{split}
&\frac{\partial c}{\partial t}=D\Delta \mu,\\
&\mu=c^3-c-\gamma\Delta c, \\
\end{split}
\end{flalign}
where $c\in[-1,1]$ is the phase concentration, $\mu$ is the chemical potential, $\sqrt{\gamma}$ is the thickness of the transition region between the two phases and $D$ is a diffusion coefficient. In this particular case the solution of system \equref{eq:cahnhilliard} is investigated over the two-dimensional curved surface of a sphere in three dimensions.}

\revised{The discretization process is similar to that of \textit{example 1a}. Since the spherical geometry is generated in Cartesian coordinates, the domain itself has to be three-dimensional, however, the smoothing kernel function in the SPH interpolant should be two-dimensional due to the fact, that the problem is interpreted on a two-dimensional curved surface. Employing the second differential operator \equref{eq:diffop3}, the discretized form of the system \equref{eq:cahnhilliard} can be written as
\begin{flalign} \label{eq:cahnhilliard_sph_discretized}
\begin{split}
&\frac{\partial c_i}{\partial t}=D\sum_j{2(\mu_j-\mu_i)\frac{\textbf{r}_j-\textbf{r}_i}{\vert \textbf{r}_j-\textbf{r}_i \vert^2}\frac{m_j}{\rho_j}\nabla W_{ij}},\\
&\mu_i=c^3-c-\gamma\sum_j{2(c_j-c_i)\frac{\textbf{r}_j-\textbf{r}_i}{\vert \textbf{r}_j-\textbf{r}_i \vert^2}\frac{m_j}{\rho_j}\nabla W_{ij}}, \\
\end{split}
\end{flalign}
where again, the smoothing kernel $W_{ij}$ is normalized in two dimensions. Following the same procedure, symbols have to be slightly changed with respect to the current system of equations \equref{eq:cahnhilliard_sph_discretized}:}
\begin{lstlisting}
constants:
  - rho0: 1000
  - dx: 0.07
  - h: dx*1.05
  - mass: dx^2*rho0
  - D: 0.003
  - gamma: (2*h/3)^2
  - dt_g: gamma/(2*h)^2
variables:
  - dt: 0.003
  - Time: 0
  - print_interval: dt
fields:
  - c: rand(-1,1)
  - c_dot: 0
  - mu: 0
\end{lstlisting}
\revised{Here, the initial concentration of phase \texttt{c} is generated as a uniform random distribution within the range $[-1,1]$.

In respect of the particle system, the domain has to be extended to three dimensions, furthermore, since the spherical geometry is more complicated than the uniform grid in \textit{example 1a}, the particle positions are now imported from the external file \texttt{points.txt} by simply referring to its name in the \texttt{grid} block:}
\begin{lstlisting}
domain:
  cell_size: 2*h|2*h|2*h
  minimum: -15|-15|-15
  maximum: 15|15|15
  boundary: 0|0|0
grid:
  gid: 0
  file: points.txt
\end{lstlisting}
\revised{Finally, the user-defined equations are constructed using the second order \texttt{sph\_L0} operator and the same Wendland smoothing kernel function}
\begin{lstlisting}
equations:
  - eq1: Time=Time+dt # measuring elapsed time
  - eq2: mu=c^3-c-gamma*sph_L0(c,mass,rho0,Wp52220,2*h)
  - eq3: c_dot=D*sph_L0(mu,mass,rho0,Wp52220,2*h)
  - eq4: c=euler(c,c_dot,dt)
  - eq5: dt=0.1*min(1/fmax(c_dot),dt_g) # adaptive time stepping
  - eq6: print_interval=exp(Time/45)
\end{lstlisting}
\revised{Here, \texttt{sph\_L0} operator implements the Laplacian \equref{eq:diffop3}. Since the phase separation process has a decaying intensity in the function of time, the \texttt{print\_interval} is adaptively modified with an exponential law in \texttt{eq6} to minimize the amount of simulation result files and keep the temporal sampling appropriate at the same time. The evolution of phase separation is visualized in Figure \ref{fig:cahnhilliard_result} by Delaunay-triangulation of the particle system in Paraview.}
\customfigure{0.6}{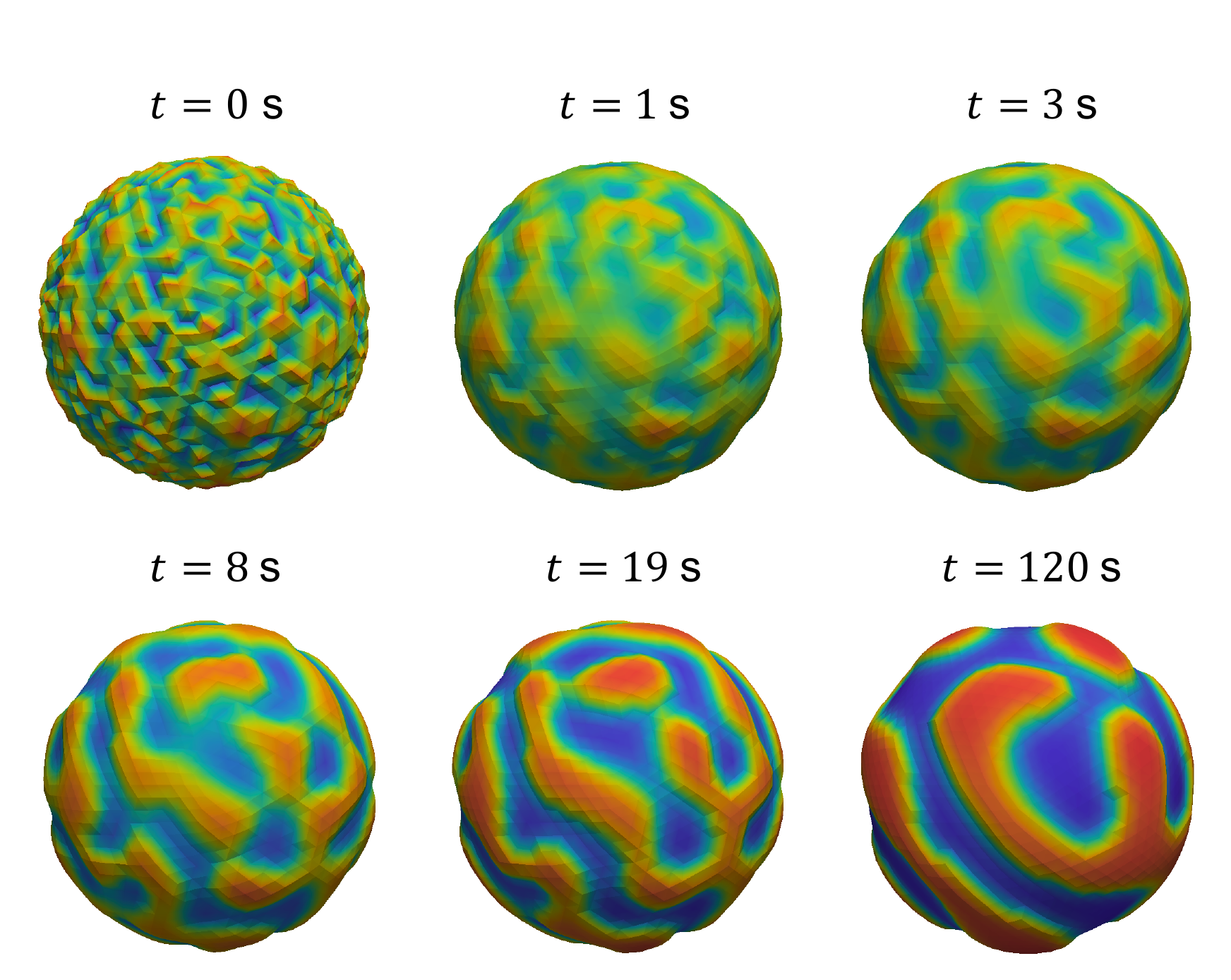}{Evolution of phase separation on the sphere.}

\subsection{Example 2 - Simulation of a particle damper (DEM)}
Particle dampers are one of the widely investigated passive damper systems \revised{today}. Although there exist several analytical models like \cite{Olson2003} or \cite{Saeki2005} to investigate and design a particle damper, the complexity of the problem still requires experimental and numerical investigation. The geometry of the tank, the number and size of particles, materials, the operating frequency are only some of the huge amount of possibilities concerning the development of particle dampers.
\subsubsection{Problem definition}
Consider a simple three-dimensional oscillating cubic tank filled with spheres of identical radii. The tank is initially at rest in the position $z_0=-0.05$ m. The layout of the particle damper is presented in Figure \ref{fig:particle_damper_geom}, furthermore the values of the introduced quantities are summarized in Table \ref{tbl:particle_damper_values}. The system is supported by an ideal linear spring merely damped by the collision of the included set of spheres.
\customfigure{0.4}{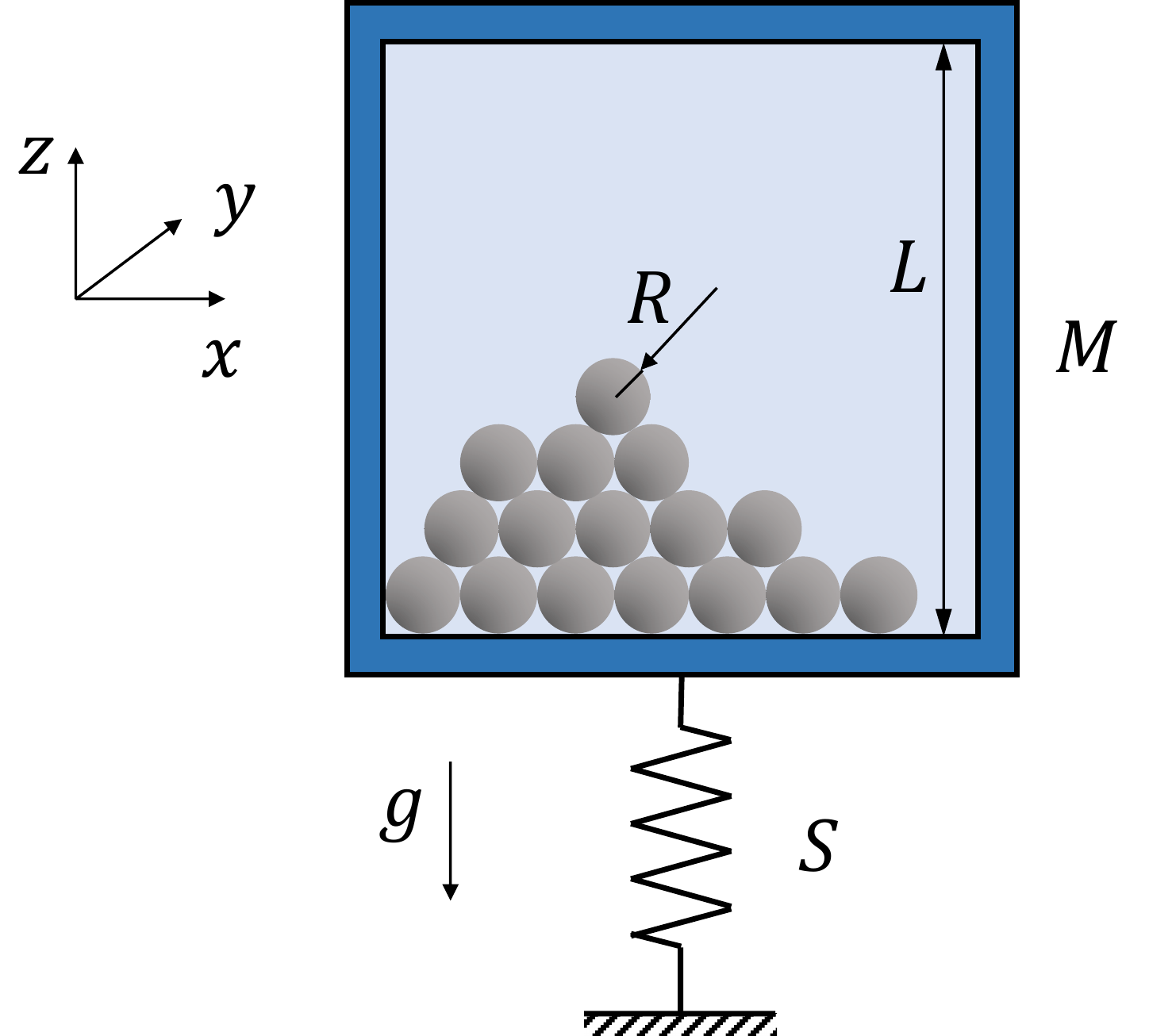}{Physical layout of the particle damper.} 
\begin{table}[H]
\begin{center}
\caption{Parameters of the particle damper simulation.}\label{tbl:particle_damper_values}
\begin{tabular}{ c l r } 
\toprule[1.5pt]
\bf Name & \bf Description & \bf Value \\
\midrule
$M$ & Tank mass & $20$ kg \\
$S$ & Spring stiffness & $78956.8$ kg/s$^2$ \\
$R$ & Particle radius & $4$ mm \\
$L$ & Tank edge length & $0.1$ m \\
$\rho$ & Particle mass density & $7850$ kg/m$^3$ \\
$E$ & Particle Young modulus & $2.06$ MPa \\
$\nu$ & Particle Poisson's ratio & $0.33$ \\
$g$ & Gravitational acceleration & $-9.81$ m/s${^2}$ \\
$N$ & Number of particles & 567 \\
\bottomrule[1.25pt]
\end{tabular}
\end{center}
\end{table}
The one-dimensional equation of motion of the tank is
\begin{flalign} \label{eq:tank_ode}
M\ddot{Z}+SZ=-\textbf{F}\textbf{e}_z,
\end{flalign}
\begin{flalign} \label{eq:boundary_force}
\textbf{F} = \sum_i{\textbf{F}^b_i}
\end{flalign}
\revised{where $Z$ is the vertical position of the tank, $\textbf{e}_z$ is the unit vector in $z$-direction}, $\textbf{F}^b$ is the resultant of the particle-boundary forces $\textbf{F}^b_i$ appearing in the equation of motion of the particles:
\begin{flalign} \label{eq:spheres_ode}
\frac{d^2\textbf{r}_i}{dt^2}=\frac{\textbf{F}^c_i}{m_i}+\frac{\textbf{F}^b_i}{m_i}+\textbf{g},
\end{flalign}
where $\textbf{F}^c$ is the particle-particle collision force and $\textbf{g}=-9.81\textbf{e}_z$ is the gravitational acceleration.

\subsubsection{Numerical model and results}
The motion of large number of colliding individual particles is often simulated using the Discrete Element Method (DEM) directly calculating the interparticle collisions based on different contact models. 

A simple representation of the three-dimensional mathematical \revised{model} is introduced in this section with the notation that other valid solutions are also possible. To simplify the model and omit the tank the simulation domain is chosen to be the interior of the tank. Since the domain is fixed, this assumption means that the simulation of the particle motion and collision is interpreted in the moving coordinate system associated to the tank and the excitation of the particles is governed purely by a time-dependent acceleration field superposed with the gravitational acceleration. The boundaries of the domain are set to be symmetric, which plays an important role in the calculation of the forces acting on the tank. For the sake of simplicity, the angular momentum of the particles is neglected. The solution of the homogeneous part of \equref{eq:tank_ode} is the harmonic function
\begin{flalign} \label{eq:tank_sol}
Z(t)=C_1sin(\gamma t)+C_2cos(\gamma t),
\end{flalign}
where $C_1$ and $C_2$ are constants depending on the initial conditions and $\gamma^2=S/M$. Due to the lack of damping, the oscillation yet has constant amplitude. The particles' motion is determined by \equref{eq:spheres_ode}:
\begin{flalign} \label{eq:spheres_ode_full}
\frac{d^2\textbf{r}_i}{dt^2}=\frac{\textbf{F}^c_i}{m_i}+\frac{\textbf{F}^b_i}{m_i}+\textbf{g}=\frac{1}{m_i}\sum_j{\textbf{f}^c_{j}(\textbf{r}_{ji},v_{ji},...)}+\frac{\textbf{F}^b_i}{m_i}+\textbf{g},
\end{flalign}
where $\textbf{f}^c_{j}$ is the sum of the normal and tangential interparticle forces based on the Hertzian contact model \cite{Tsuji1992} \revised{and neglecting the angular velocities and their influence on the particle motion:}
\begin{equation} \label{eq:DEM_force}
\begin{split}
\textbf{f}^c_{j}&=\textbf{f}^n_{j}+\textbf{f}^t_{j}=\big(k_{Hz}\delta^{3/2}+c_{Hz}\delta^{1/4}\dot{\delta}\big)\textbf{n}_{ji}+c_f\vert\textbf{f}^n_{j}\vert\frac{\textbf{v}^t}{\vert\textbf{v}^t\vert}, \\
\textbf{n}_{ji}&=\frac{\textbf{r}_j-\textbf{r}_j}{\vert\textbf{r}_j-\textbf{r}_j\vert},\\
\textbf{v}^t&=-\textbf{v}_{ij}+(\textbf{v}_{ij}\textbf{n}_{ji})\textbf{n}_{ji}, \\
k_{Hz} &= \frac{4}{3}\sqrt{R'}E', \\
c_{Hz} &= \frac{\sqrt{m'k_{Hz}}}{8}.
\end{split}
\end{equation}
\revised{Here $c_f$ is the coefficient of interparticle Coulomb-friction.} Since the deformation of the spheres is neglected, the contact theory models the interparticle forces as functions of the particle-overlap $\delta=R_i+R_j-\vert\textbf{r}_j-\textbf{r}_i\vert$. The effective quantities are expressed as
\begin{align} \label{eq:effective_quantities}
\begin{split}
&R'=\frac{R_iR_j}{R_i+R_j}, \\
&m'=\frac{m_im_j}{m_i+m_j}, \\
&E'=\frac{E_iE_j}{E_j(1-\nu_i^2)+E_i(1-\nu_j^2)}.
\end{split}
\end{align}
\revised{In Nauticle, the model described by equations \equref{eq:DEM_force}-\equref{eq:effective_quantities} is implemented under the operator name \texttt{dem\_l}. The second term including $\textbf{F}^b_i$} on the RHS of \equref{eq:spheres_ode} operates with the same collision laws at the symmetric boundaries, which in turn contributes to \equref{eq:tank_ode}. During the simulation, the tank position, velocity and acceleration has to be calculated at each time steps. These quantities are considered as variables and calculated by the numerical solution of \equref{eq:tank_ode}.

After running the \revised{simulation} in Nauticle, the individual particle elevations are visualized in Figure \ref{fig:particle_damper_time_series} together with the bottom and top positions of the tank.
\customfigure{0.5}{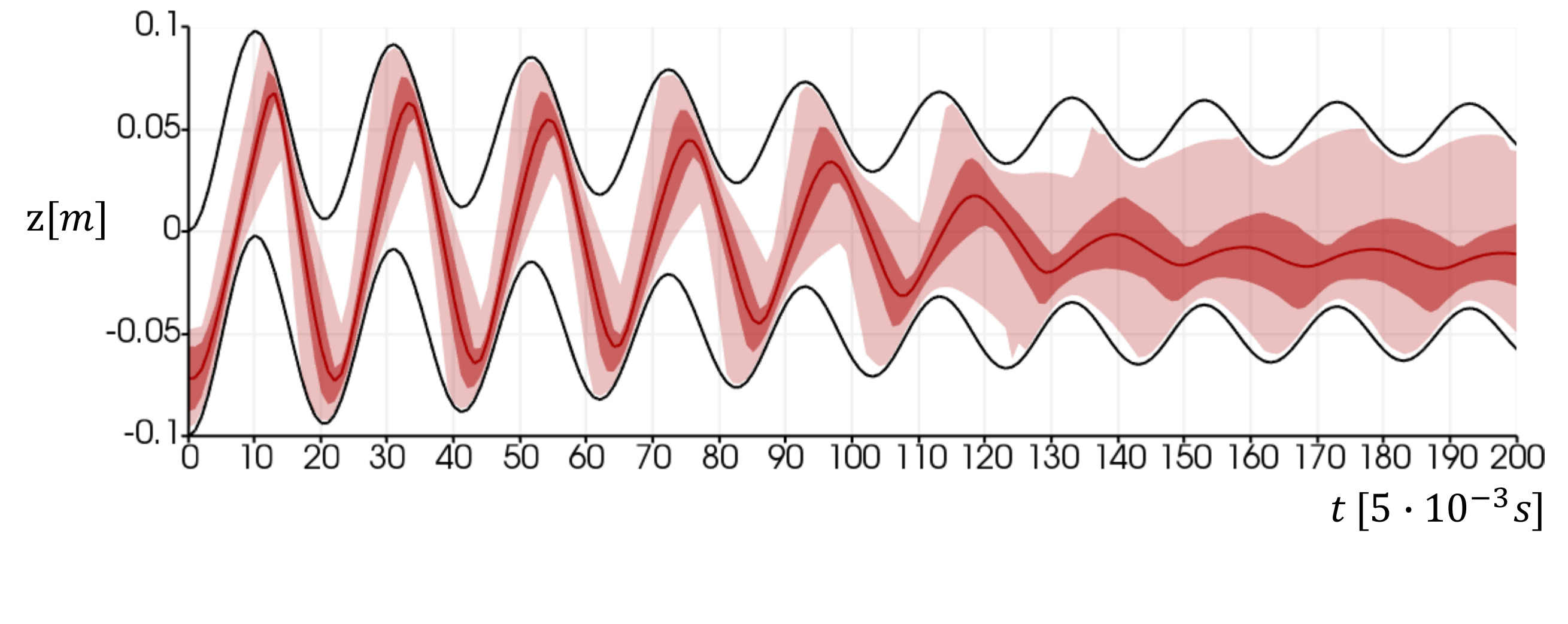}{Elevation of particles (red) and the tank (black) in the function of time.}
As it can be seen the oscillation amplitude is being reduced significantly until the particles start to gather at the bottom due to the decaying peak acceleration.

\section{\revised{Implementation of an interaction law}} \label{sec:implementation_of_interaction_law}
\revised{This section introduces the main steps of the implementation of particle interaction laws on the second, intermediate level of Nauticle. To guide the reader through the implementation process, a basic social force model has been chosen as an interaction law, which is, on the one hand, sufficiently simple, while on the other hand, suitable for the demonstration of generality and flexibility. Note that the implementation below slightly differs from the one in the current release for better clarity and readability. Although the implementation is simple and straightforward, this paper focuses on the concept and the main implementation issues omitting subsidiary operations.}
\subsection{\revised{A basic social force model}}
Evacuation time and efficiency acquire crucial importance during the design process of modern buildings. The demand \revised{for} safety protocols facilitates the research of crowd motion under predefined conditions. During the recent decades, several models of different fundamentals were built to simulate the flow of people in buildings of complex geometries based on the fundamental work of D. Helbing and P. Moln\'ar \cite{Helbing1995}. More recent models like \cite{Tissera2012} (analogy with fluid mechanics) or \cite{Li2015} (implying Cellular Automata (CA)) were built to simulate large-scale dynamics of pedestrians.
The standard \revised{social force model} is a micro-scale deterministic model considering the intentions of each person as driving forces besides the repulsive forces during collisions of their bodies. \revised{According to the model presented in \cite{HelbingFV2000}, the current implementation uses the governing equation for each individual $i$}
\begin{flalign} \label{eq:social_force_model}
&\frac{d\textbf{v}_i}{dt}=\frac{v_0\textbf{e}_0-\textbf{v}_i}{\tau}-\frac{1}{m_i}\sum_{j\neq i}{\bigg[A_i \exp\bigg(\frac{R_{ij}-d_{ij}}{B_i}\bigg)+k(R_{ij}-d_{ij})-c_{ij}\bigg]\textbf{n}_{ji}},
\end{flalign}
where $m_i$ and $\textbf{v}_i$ are the mass and velocity of the $i$th person respectively, $v_0$ is the desired velocity magnitude in the direction $\textbf{e}_0$ and $\tau$ is the time scale. Furthermore, $A_i$, $B_i$, $k$ are constants of repulsive, $c_{ij}$ is of attractive forces, $R_{ij}=R_i+R_j$ is the sum of the radii of the two individuals in collision. $\textbf{n}_{ji}$, as formerly, is the normalized direction vector pointing from person $i$ to $j$, and finally, $d_{ij}$ is the distance between them. The desired velocity vector $v_0\textbf{e}_0$ is continually changing as the person moves towards the desired position.

\subsection{\revised{Extension of SFL}}
\revised{Since numerical models in Nauticle are defined solely through the SFL (cf. Section \ref{sec:class_hierarchy_of_expressions}), new capabilities, hence interaction laws can be implemented by merely extending the SFL-tree. The recursive evaluation of user-defined expressions (cf. Figure \ref{fig:recursive_execution}) is performed by the \texttt{evaluate(i, level=0)} member function implemented in each of the non-abstract classes in the SFL-tree, where \texttt{i} and \texttt{level} are the particle index and the optional (zero by default) time level of evaluation respectively. The latter plays role only in case of multistep integrators and ignored in the present paper for the sake of simplicity.

The aim of the extension of SFL is that the users should be able to construct and apply their own interaction laws flexibly without more profound modifications of the code. To fulfill this requirement, the abstract interaction class (interface) in Figure \ref{fig:class_hierarchy} performs all preliminary operations (e.g., selection of particle pairs) and provides necessary data required for computation of pairwise interactions. In the subsequent paragraphs, it will be shown how the pairwise computations are separated from the code at the intermediate level.

Besides a few minor tasks discussed in the user's guide, users need to define a new class inherited (cf. Listing \ref{list:user_defined_class}) from the interaction class and implement the evaluation member function focusing on pairwise calculations.}
\begin{lstlisting} [style=CPP, caption={\revised{Declaration of interaction class}}, label={list:user_defined_class}]
class Social_interact : public pmInteraction<10> {
public:
  /*...*/ // constructors, destructor, etc.
  pmTensor evaluate(int const& i, 
                    size_t const& level=0) const override;
};
\end{lstlisting}
\revised{The integer \texttt{10} (it can be any non-negative integer) as a template argument of the \texttt{pmInteraction} class defines the number of parameters of the current interaction law. In order to prevent accidental multithreading race conditions on the lowest level, the \texttt{const} keyword (beside \texttt{override}) is also needed at the declaration assuring that neither objects instantiated from the classes of the SFL-tree can be modified through the evaluation function.

The definition of the \texttt{evaluate} member function should consist of three parts. Firstly, in this particular case, the ten operands (which is an exceptionally large number compared to other interactions) of the interaction need to be acquired and stored now in an array \texttt{pmi}:}
\begin{lstlisting} [style=CPP, caption={\revised{Storing operands (inside evaluate function)}}, label={list:evaluation_function_part1}]
double cell_size_min = this->psys.lock()->\
  get_particle_space()->get_domain().get_cell_size().min();
double pmi[10];
for(int i=0;i<10;i++) {
  pmTensor pmi[i] = this->operand[i]->evaluate(i,level);
}
\end{lstlisting}
\revised{The \texttt{cell\_size\_min} variable is obtained from the particle system $psys$ object assigned to the parent pmInteraction class (cf. Section \ref{sec:class_hierarchy_of_expressions}). Note that each of the operands is being evaluated through their own \texttt{evaluate} functions, leading to the possibility of nested operators in SFL. Secondly, as the most important part, the calculation of the pairwise interaction should be defined concerning particle $i$ and $j$ based on equation \equref{eq:social_force_model}:}
\begin{lstlisting} [style=CPP, caption={\revised{Lambda function expressing the interaction between particles \texttt{i} and \texttt{j} (inside evaluate function)}}, label={list:evaluation_function_part2}]
auto contribute = [&](pmTensor const& rel_pos,
                      int const& i, int const& j,
                      pmTensor const& cell_size,
                      pmTensor const& guide)->pmTensor{
  double Rj = this->operand[4]->evaluate(j,level)[0];
  double cj = this->operand[8]->evaluate(j,level)[0];
  pmTensor contribution{2,1,0.0};
  double d_ji = rel_pos.norm();
  if(d_ji > 0 && d_ji < cell_size_min) {
    pmTensor n_ji = rel_pos/d_ji;
    double Rij = pmi[2]+Rj;
    double cij = (pmi[6]+cj)*0.5;
    double body_force = d_ji-Rij<0.0 ? pmi[5]*(Rij-d_ji) : 0.0;
    contribution = -pmi[3]*std::exp((Rij-d_ji)/pmi[4])*n_ji;
    contribution -= body_force*n_ji - cij*n_ji;
    contribution /= pmi[1];
  }
  return contribution;
};
\end{lstlisting}
\revised{Finally, the lambda (nested function) should be passed to the \texttt{interact} function inherited from the \texttt{pmInteraction} class and the result can be returned:}
\begin{lstlisting} [style=CPP, caption={\revised{Return the sum of contributions (inside evaluate function)}}, label={list:evaluation_function_part3}]
pmTensor e0 = (pmi[1]-this->psys.lock()->get_value(i));
e0 /= e0.norm();
return (pmi[0]*e0-pmi[0])/pmi[7] + this->interact(i, contribute);
\end{lstlisting}
\revised{The complete listing of the evaluate function is shown in \ref{app:code_social_force}. At the intermediate developer level, the \texttt{interact(i, contribute)} term can be considered as a black box summing up all the contributions to particle \texttt{i}.}

\subsection{\revised{Simulation based on the social force model}}
\revised{As a demonstration of the implemented model, a simulation result of the evacuation of a dummy building is qualitatively presented here.} The constants $A$, $B$, $k$, $c$ \revised{in \equref{eq:social_force_model}} are considered to be identical for each individual person. The desired position of each person is calculated based on the current \revised{positions}, hence it is updated as a person leaves a room. The mass of the individuals is randomly distributed between $50$ kg and $100$ kg and used to calculate their sizes (radii) with the linear function
\begin{flalign}
R_i=0.002m_i+0.15.
\end{flalign}
\revised{Besides the initial positions, the} desired velocities are also randomly chosen independently from \revised{other properties of the individuals}. To prevent people from crossing any of the walls, the boundary conditions -- hence the building itself -- is built up using particles having the same properties as \revised{individual} people except that they are fixed in space during the whole simulation. The $R_w$ radii of the wall-particles are constant and $R_w=0.1$ m. \revised{The snapshots of the process in different time instants are shown by Figure \ref{fig:evacuation_result}.}
\customfigure{0.6}{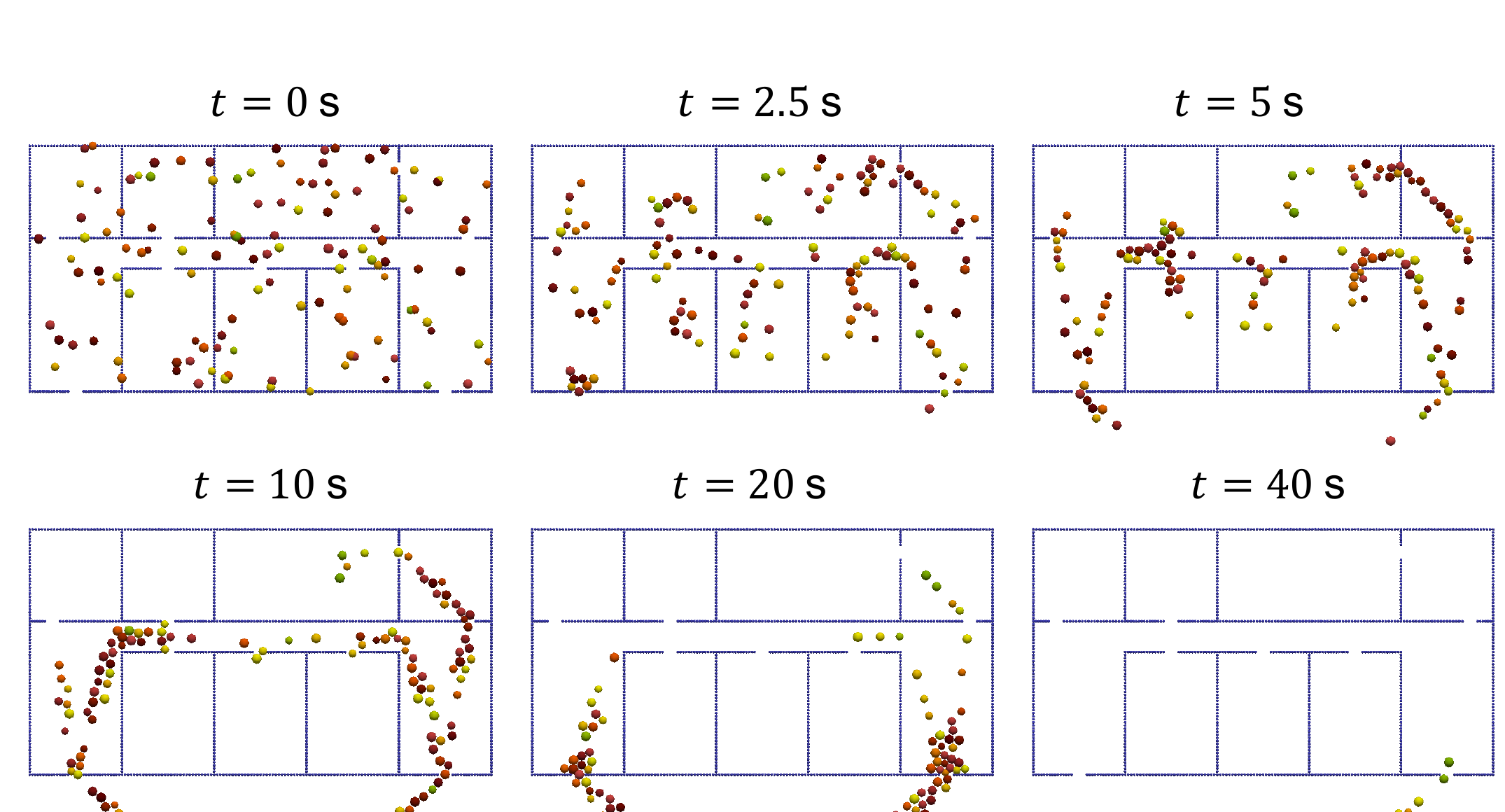}{Evacuation process based on the SFM-model. The size of the particles is equal to their radii and colored by the associated desired velocities.}

\section{\revised{Code efficiency, performance benchmark}} \label{sec:performance_benchmark}
\revised{Achieving reasonable computational speed while maintaining freedom for the configuration of mathematical models is a challenging task. As one of the possible implementations, using the Domain Specific Language (DSL) provided by the highest abstraction level, Aboria constructs the mathematical model on a low level at compile time resulting in lossless computational efficiency compared to a hard-coded solution \cite{Aboria}. Contrastingly, the SFL introduced in Section \ref{sec:class_hierarchy_of_expressions} makes the interpretation and solution at runtime without precompilation of the user-defined equations. Although the built-in interaction operators of Nauticle are compiled, the evaluation in interpreter mode implies a certain computational overhead in return for self-explanatory configuration by SFL.

For comparison of computational performance, a simple test has been identically built with both Aboria and Nauticle implying a three-dimensional sampling of a scalar field using the SPH method, which is -- according to a large number of neighbors $n\approx39$ -- one of the most computationally demanding operators of interactions. The domain is chosen to be cubic with periodic boundaries on each side and filled with particles placed along an axis aligned uniform rectilinear grid. The operator to sample the scalar field $A$ reads as
\begin{equation}
S_i=\sum_j^{n_i}{A_j\frac{m_j}{\rho_j}W_{ij}},
\end{equation}
where $S$ is the field storing the sampled data.
The computations have been performed with the various number of particles from $1$ to $10^5$. Figure \ref{fig:speed_comparison} shows the efficiency of the computations $N/T_e$, representing the number of processed particle per second as a function of the particle number $N$ in case of Nauticle with varying number of computational threads and Aboria. $T_e$ denotes the total evaluation time of each simulation. According to the expectations, the larger the number of particles, the simulations with more involved threads become more efficient. Throughout roughly three orders of magnitudes of $N$, Aboria performs approximately the same as Nauticle with four simultaneous threads, while the twelve-threaded simulations dominate above $N=10^2$, with the peak of $N/T_e=1.05\cdot10^3 s^{-1}$ at $N=3375$.

Although the arrears in the speed of Nauticle is significant (compared to Aboria), it performs well in multithreaded runs and has the potential to achieve higher performance by the precompilation of the currently runtime-evaluated SFL-structure.}

\customfigure{0.5}{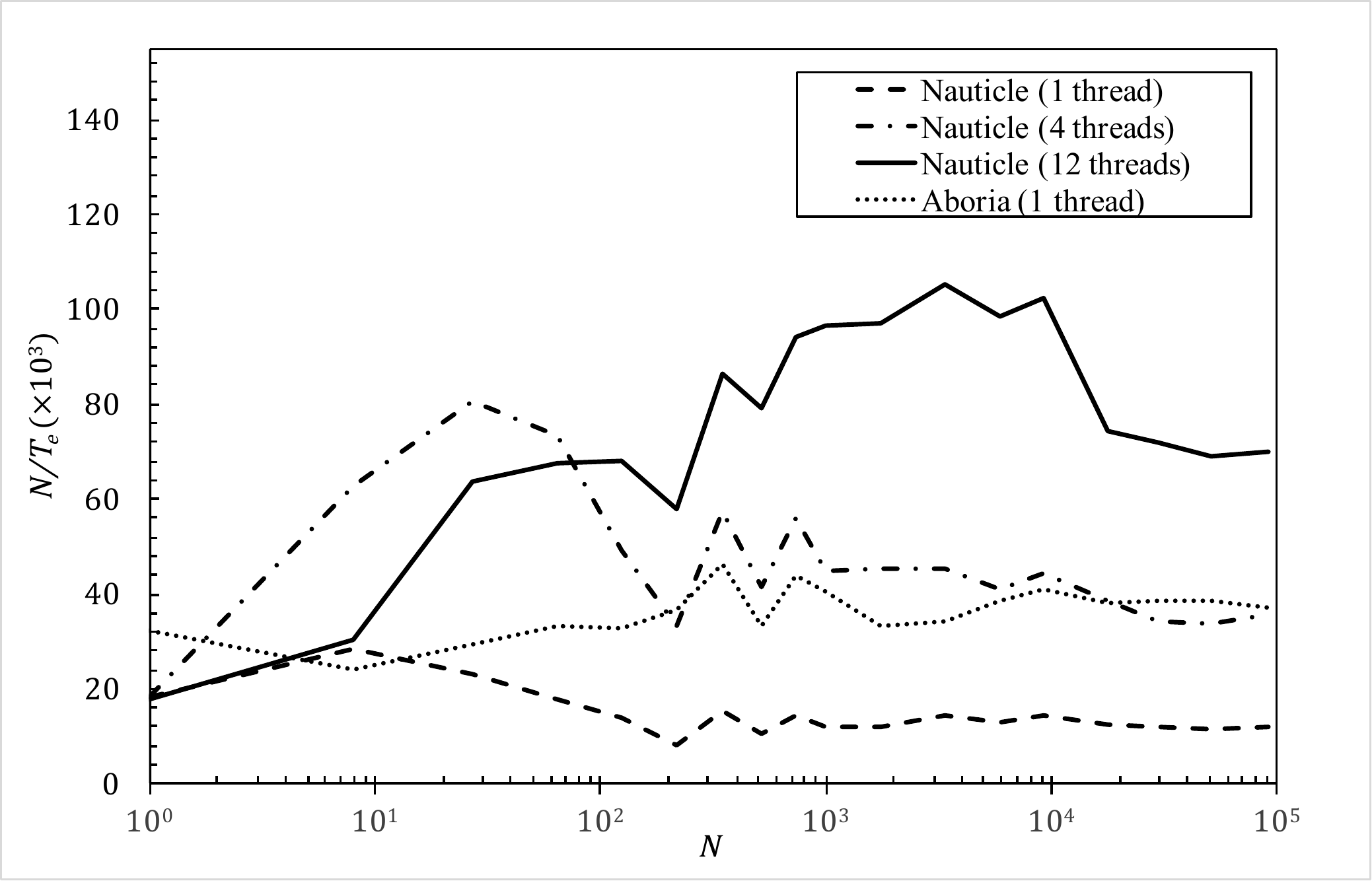}{\revised{Computational efficiency of Nauticle and Aboria. The number of processed particles per second $N/T_e$ plotted as a function of all particles $N$.}}

\section{Future \revised{improvements}} \label{sec:future_improvements}
The development of Nauticle is started around the second half of 2015. Initially, the code was built up using former particle-based algorithms written by the author. \revised{Following the basic ideas aiming generality, after} the implementation of the solver core \revised{including the SFL environment} the software became a useful simulation tool in several research areas. However, being a small and new project, further developments are required to extend the capabilities of the solver.

\revised{Some of the most important} features planned to be included in the \revised{further} versions of Nauticle \revised{are the following:}
\begin{enumerate}
\item Generation and runtime compilation of user-defined equations to increase computational efficiency.
\item Extend the environment for implicit meshless schemes and other meshless interpolants, e.g., Moving Least Squares (MLS) interpolant.
\item Implementation of particle sources and sinks.
\end{enumerate}

\section{\revised{Summary and conclusions}} \label{sec:conclusions}
The present paper introduces the novel general-purpose meshless particle-based numerical simulation tool Nauticle, which facilitates both the \revised{flexible} application and \revised{development} of meshless particle-methods. \revised{It is stated that the Nauticle code has three distinct levels for both users and developers, then the top two layers are presented and explained through simple examples.}

As a fundamental concept, all particle methods are considered as interaction laws between physically existing or abstract individual elements (particles) \revised{and} the general form of \revised{the} governing equations that can be solved using Nauticle \revised{is introduced}. The basic implementaion idea is to move the mathematical model up from the core of the solver to the \revised{users'} level \revised{by providing} the \revised{presented Symbolic Form Language (SFL). It is shown that using SFL, excessively intuitive and straightforward YAML configuration files can be constructed without any programming effort.}

\revised{It is also shown that -- on the intermediate developer level (second level) -- the SFL class hierarchy of expressions implemented in C++11 makes it possible to adopt new interaction laws by the simple extension of the hierarchy through the node of the Interaction interface. Although it requires C++ programming experience, the core of the code barely needs to be known for such an extension.}

The main features of the current Nauticle release are summarized below:
\begin{itemize}
  \item \revised{definition of symbols and equations using the SFL in YAML documents,}
  \item solution of \revised{arbitrary} symbolic user-defined governing equations in one, two or three dimensions,
  \item periodic and symmetric boundary conditions in the computational domain as an axis-aligned rectangular box,
  \item hot start simulations using former results as initial conditions written in binary or ASCII VTK files,
  \item adoption of new particle schemes by merely writing an interaction class derived from the interaction node of the expression tree.
\end{itemize}

\revised{In consideration of the listed features, Nauticle is exceptionally suitable not only for the application of particle methods but also for testing of new theories, mathematical models or even geometries. Although, coupling problems are out of the scope of the present paper, the arbitrary combination of particle methods is also trivially feasible through the SFL.}

\appendix
\section{ Configuration of example 1a} \label{app:config_example1a}
\begin{lstlisting}
simulation:
  case:
    workspace:
      constants:
        - rho0: 1000
        - h: 0.25
        - dx: h/1.1
        - mass: dx^2*rho0
        - c: 50
        - g: 0|-9.81
      variables:
        - dt: 1e-3
      particle_system:
        domain:
          cell_size: 2*h|2*h
          minimum: 0|0
          maximum: 7/h|10/h
          boundary: symmetric|symmetric
        grid:
          gid: 0
          gpos: 0|0
          gsize: 7|4
          goffset: 0|0
          gip_dist: dx|dx
      fields:
        - rho: rho0
        - rhodot: 0
        - v: 0|0
        - vdot: 0|0
        - p: 0
    equations:
      - eq1: rhodot=-rho*sph_D00(v,mass,rho,Wp52220,2*h)
      - eq2: rho=euler(rho,rhodot,dt)
      - eq3: p=c^2*(rho-rho0)
      - eq4: vdot=-1/rho*sph_G11(p,mass,rho,Wp52220,2*h)
      - eq5: vdot=vdot+0.1*c*h*sph_A(v,mass,rho,Wp52220,2*h)+g
      - eq6: v=euler(v,vdot,dt)
      - eq7: r=euler(r,v,dt)
  parameter_space:
    simulated_time: 6
    print_interval: 0.05

\end{lstlisting}
\section{ Configuration of example 1b} \label{app:config_example1b}
\begin{lstlisting}
simulation:
  case:
    workspace:
      constants:
        - rho0: 1000
        - dx: 0.07
        - h: dx*1.05
        - mass: dx^2*rho0
        - D: 0.003
        - gamma: (2*h/3)^2
        - dt_g: gamma/(2*h)^2
      variables:
        - dt: 0.003
        - Time: 0
        - print_interval: dt
      particle_system:
        domain:
          cell_size: 2*h|2*h|2*h
          minimum: -15|-15|-15
          maximum: 15|15|15
          boundary: 0|0|0
        grid:
          gid: 0
          file: points.txt
      fields:
        - c: rand(-1,1)
        - c_dot: 0
        - mu: 0
    equations:
      - eq1: Time=Time+dt
      - eq2: mu=c^3-c-gamma*sph_L0(c,mass,rho0,Wp52220,2*h)
      - eq3: c_dot=D*sph_L0(mu,mass,rho0,Wp52220,2*h)
      - eq4: c=euler(c,c_dot,dt)
      - eq5: dt=0.1*min(1/fmax(c_dot),dt_g)
      - eq6: print_interval=exp(Time/45)
  parameter_space:
    simulated_time: 150
    print_interval: 0.001
\end{lstlisting}
\section{ Implementation of social force interaction} \label{app:code_social_force}
\begin{lstlisting} [style=CPP]
pmTensor Social_interact::evaluate( int const& i,
                                    size_t const& level=0) const
{
  double cell_size_min = this->psys.lock()->\
    get_particle_space()->get_domain().get_cell_size().min();
  double pmi[10];
  for(int i=0;i<10;i++) {
    pmTensor pmi[i] = this->operand[i]->evaluate(i,level);
  }
  auto contribute = [&](pmTensor const& rel_pos,
                        int const& i, int const& j,
                        pmTensor const& cell_size,
                        pmTensor const& guide)->pmTensor{
    double Rj = this->operand[4]->evaluate(j,level)[0];
    double cj = this->operand[8]->evaluate(j,level)[0];
    pmTensor contribution{2,1,0.0};
    double d_ji = rel_pos.norm();
    if(d_ji > 0 && d_ji < cell_size_min) {
      pmTensor n_ji = rel_pos/d_ji;
      double Rij = pmi[2]+Rj;
      double cij = (pmi[6]+cj)*0.5;
      double body_force = d_ji-Rij<0.0 ? pmi[5]*(Rij-d_ji) : 0.0;
      contribution = -pmi[3]*std::exp((Rij-d_ji)/pmi[4])*n_ji;
      contribution -= body_force*n_ji - cij*n_ji;
      contribution /= pmi[1];
    }
    return contribution;
  };
  pmTensor e0 = (pmi[1]-this->psys.lock()->get_value(i));
  e0 /= e0.norm();
  return (pmi[0]*e0-pmi[0])/pmi[7] + this->interact(i,contribute);
}
\end{lstlisting}

\newpage
\bibliographystyle{elsarticle-num-notitle}
\bibliography{references}

\end{document}